\documentstyle[preprint,pra,aps,eqsecnum,amsfonts]{revtex}

\tightenlines

\begin{document}

\draft
\title{Asymptotics of Clebsch-Gordan Coefficients}
\author{Matthias W. Reinsch and James J. Morehead}
\address{Department of Physics, University of California, Berkeley, California
94720}
\date{\today}
\maketitle

\begin{abstract}

Asymptotic expressions for Clebsch-Gordan coefficients are derived from
an exact integral representation.  Both the classically allowed and forbidden
regions are analyzed.  Higher-order approximations are calculated.  These
give, for example, six digit accuracy when the quantum numbers are
in the hundreds.

\end{abstract}


\newpage

\section{Introduction}

This paper contains a detailed study of the asymptotics of
Clebsch-Gordan coefficients and includes the derivation of new
results.  We use the term ``Clebsch-Gordan coefficient'' in its
colloquial sense, {\it i.\ e.\ }the vector addition coefficients of
$SU(2)$.  Thus our results also give the asymptotics of the
$3j$-symbols.  We consider the case in which all of the quantum
numbers get large together.  What this means is multiplying all of the
quantum numbers by a number and studying the asymptotic behavior of
the Clebsch-Gordan coefficient as this multiplier gets large.  Such a
multiplier is often called $1/\hbar$, so that the limit of large
quantum numbers is the limit of small $\hbar$.

The history of this subject dates back to the early days of quantum
mechanics and the study of the classical limit of quantum mechanical
quantities.  Numerous papers have been written in this area.  We
summarize the literature briefly here.  In 1959, Wigner\cite{Wigner}
discussed the physical interpretation and classical limits of
Clebsch-Gordan coefficients.  He described a certain average behavior,
and did not analyze the oscillatory nature of the Clebsch-Gordan
coefficients.  There are references in this work to
Edmonds\cite{edmonds} and Brussaard and Tolhoek\cite{bandt}.  In 1968,
Ponzano and Regge\cite{PR} presented asymptotic expressions that
included the oscillations.  Their work included an interpretation of
certain angles that occur in their results and in ours.  Additionally,
they discussed the allowed and forbidden regions.  However their
derivation is, in their words, ``rather heuristic.''  It was borne out
in their comparisons with the exact values.  William Miller\cite{BM}
derived similar expressions using semiclassical methods in 1974, but
did not treat the forbidden region.  Another work that relates to the
present paper is that of Srinivasa Rao and V. Rajeswari\cite{RR}. It
contains exact expressions for Clebsch-Gordan coefficients and their
relationship to certain hypergeometric series.  There is more
information in the work of Biedenharn and Louck\cite{BL}.

In this paper, we start by deriving an exact integral representation
for the Clebsch-Gordan coefficients.  Then the methods of stationary
phase are used to approximate this integral.  The allowed and
forbidden regions are treated separately, and the resulting
expressions are related to the literature.  These methods are then
used to derive higher-order results, that is, the next order in an
expansion in $\hbar$.  These formulas are accurate to five or six
digits when the quantum numbers are in the hundreds.

Possible applications of this work include high-angular momentum
calculations and theoretical investigations which contain sums over
large numbers of Clebsch-Gordan
coefficients\cite{generatingfunctions,Labarthe}.

\section{Exact expressions for the Clebsch-Gordan coefficient}

Our starting point is an exact expression for
the Clebsch-Gordan (vector-addition) coefficient, due to Wigner
(see, for example, Eq.~(3.6.11) of Ref.~\onlinecite{edmonds}),
\begin{displaymath}
\langle j_1\,m_1\,j_2\,m_2 \,|\,j\,m \rangle =
\left[
(2j+1)(j_1\!+\!j_2\!-\!j)!(j_1\!-\!j_2\!+\!j)!(-j_1\!+\!j_2\!+\!j)! /
(j_1\!+\!j_2\!+\!j\!+\!1)!\right] ^{1/2}
\end{displaymath}
\begin{eqnarray}
&\times&
[(j_1\!+\!m_1)!(j_1\!-\!m_1)!(j_2\!+\!m_2)!(j_2\!-\!m_2)!(j\!+\!m)!
(j\!-\!m)!]^{1/2}  \nonumber \\
&\times&
\sum_z  
\frac{(-1)^z}{z!(j_1\!+\!j_2\!-\!j\!-\!z)!(j_1\!-\!m_1\!-\!z)!
(j_2\!+\!m_2\!-\!z)!(j\!-\!j_2\!+\!m_1\!+\!z)!(j\!-\!j_1\!-\!m_2\!+\!z)!
}\;.
\label{Wigner}
\end{eqnarray}
A factor of $\delta_{m,m_1+m_2}$ has been omitted; throughout this
paper we will assume that $m$ is equal to $m_1+m_2$.  Also, unless
otherwise specified, sums over an index are sums over all integers.  It will
turn out, though, that the summand is nonzero for only finitely many
values of the index.

We begin by deriving the following exact expression for the Clebsch-Gordan
coefficient.
\begin{displaymath}
\langle j_1\,m_1\,j_2\,m_2 \,|\,j\,m \rangle =
(-1)^{j+m} \, N_{j_1 \, m_1 \, j_2 \, m_2 \, j \, m} \,
\frac{1}{(j_1-m_1)! \, (j_2-m_2)!}
\end{displaymath}
\begin{equation}
\times
\left(\frac{d}{du}\right)^{j_1-m_1} \,
\left(\frac{d}{dt}\right)^{j_2-m_2} \, \left. \left[(t-1)^{j+j_2-j_1} \,
(t-u)^{j_1+j_2-j} \, (u-1)^{j+j_1-j_2} \right] \, \right|_{u=0,t=0} \; ,
\label{eq1}
\end{equation}
where $N_{j_1 \, m_1 \, j_2 \, m_2 \, j \, m}$ is defined to be
\begin{equation}
N_{j_1 \, m_1 \, j_2 \, m_2 \, j \, m} =
\left[\frac{
(2j+1)(j_1\!+\!m_1)!(j_1\!-\!m_1)!(j_2\!+\!m_2)!(j_2\!-\!m_2)!(j\!+\!m)!
(j\!-\!m)!}{(j_1\!+\!j_2\!+\!j\!+\!1)!
(j_1\!+\!j_2\!-\!j)!(j_1\!-\!j_2\!+\!j)!(-j_1\!+\!j_2\!+\!j)!} \right]^{1/2} \; .
\label{Ndefinition}
\end{equation}
Because the quantity being differentiated in Eq.~(\ref{eq1}) is a
polynomial in the variables $u$ and $t$, the operation of
differentiating this quantity and then evaluating the result at $u=0$
and $t=0$ simply selects a particular coefficient in the polynomial.
Thus, Eq.~(\ref{eq1}) expresses the Clebsch-Gordan coefficient as a
certain coefficient in a polynomial that can be written in closed
form.  This equation can be derived from results in the
literature\cite{Regge}, but we give here an independent derivation of
Eq.~(\ref{eq1}) from Eq.~(\ref{Wigner}) to verify that all of the
conventions involved are consistent.

In order to prove Eq.~(\ref{eq1}), we start by finding the coefficient
of $u^{j_1-m_1} \, t^{j_2-m_2}$ in the polynomial $(t-1)^{j+j_2-j_1}
\, (t-u)^{j_1+j_2-j} \, (u-1)^{j+j_1-j_2}$.  This is equal to the
coefficient of $u^{j_1-m_1}$ in the polynomial that is given by
$(u-1)^{j+j_1-j_2}$ times the $u$-dependent coefficient of
$t^{j_2-m_2}$ in the polynomial $(t-1)^{j+j_2-j_1} \,
(t-u)^{j_1+j_2-j}$.  Using the Binomial Theorem, we get
\begin{equation}
(t-1)^{j+j_2-j_1} = \sum_k 
\left(\begin{array}{c} j+j_2-j_1 \\ k \end{array} \right) \,
t^k \, (-1)^{j+j_2-j_1-k}
\end{equation}
and
\begin{equation}
(t-u)^{j_1+j_2-j} = \sum_\ell 
\left(\begin{array}{c} j_1+j_2-j \\ \ell \end{array} \right) \,
t^\ell \, (-u)^{j_1+j_2-j-\ell} \, .
\end{equation}
The coefficient of $t^{j_2-m_2}$ in the product of these is
\begin{displaymath}
\sum_k \left(\begin{array}{c} j+j_2-j_1 \\ k \end{array} \right) \,
(-1)^{j+j_2-j_1-k} \,
\left(\begin{array}{c} j_1+j_2-j \\ j_2-m_2-k \end{array} \right) \,
(-u)^{j_1+j_2-j-(j_2-m_2-k)}
\end{displaymath}
\begin{equation}
= (-1)^{j_2+m_2} \, u^{j_1-j+m_2} \sum_k 
\left(\begin{array}{c} j+j_2-j_1 \\ k \end{array} \right) \,
\left(\begin{array}{c} j_1+j_2-j \\ j_2-m_2-k \end{array} \right) \,
u^k \; .
\end{equation}
As explained above, we need to multiply this polynomial by
\begin{equation}
(u-1)^{j+j_1-j_2} = \sum_\ell 
\left(\begin{array}{c} j+j_1-j_2 \\ \ell \end{array} \right) \,
u^\ell \, (-1)^{j+j_1-j_2-\ell} \, .
\end{equation}
and find the coefficient of $u^{j_1-m_1}$.  The result is
\begin{displaymath}
(-1)^{j_2+m_2} \, \sum_k 
\left(\begin{array}{c} j+j_2-j_1 \\ k \end{array} \right) \,
\left(\begin{array}{c} j_1+j_2-j \\ j_2-m_2-k \end{array} \right) \,
\left(\begin{array}{c} j+j_1-j_2 \\ j-m-k \end{array} \right) \,
(-1)^{j_1-j_2+m+k}
\end{displaymath}
\begin{equation}
= (-1)^{j+m} \,
\sum_z (-1)^z \,
\left(\begin{array}{c} j+j_2-j_1 \\ j_2+m_2-z \end{array} \right) \,
\left(\begin{array}{c} j_1+j_2-j \\ z \end{array} \right) \,
\left(\begin{array}{c} j+j_1-j_2 \\ j_1-m_1-z \end{array} \right) \, ,
\end{equation}
where we have redefined the index of summation according to $k =
j-j_1-m_2+z$ in the final line of this equation and made use of the
identity $\left(\begin{array}{c} a \\ b \end{array} \right) =
\left(\begin{array}{c} a \\ a-b \end{array} \right)$.  Since $j+m$ is
always an integer, $(-1)^{2(j+m)}$ is equal to one, and we have shown
that the right-hand side of Eq.~(\ref{eq1}) is equal to
\begin{displaymath}
N_{j_1 \, m_1 \, j_2 \, m_2 \, j \, m} \,
\sum_z (-1)^z \,
\left(\begin{array}{c} j+j_2-j_1 \\ j_2+m_2-z \end{array} \right) \,
\left(\begin{array}{c} j_1+j_2-j \\ z \end{array} \right) \,
\left(\begin{array}{c} j+j_1-j_2 \\ j_1-m_1-z \end{array} \right)
\end{displaymath}
\begin{equation}
=
\sum_z \frac{(-1)^z \, N_{j_1 \, m_1 \, j_2 \, m_2 \, j \, m} \,
(j+j_2-j_1)! \, (j_1+j_2-j)! \, (j+j_1-j_2)!}
{(j-j_1-m_2+z)! \, (j_2+m_2-z)! \,
(j_1+j_2-j-z)! \, z! \,
(j_1-m_1-z)! \, (j-j_2+m_1+z)!}
\end{equation}
This is the same as the right-hand side of Eq.~(\ref{Wigner}) and completes
the proof of Eq.~(\ref{eq1}).
An alternative proof begins by introducing a factor of $x^z$ into
the sum in Eq.~(\ref{Wigner}) and deriving a third-order differential
equation for the resulting function of $x$.  This differential
equation can be solved using hypergeometric functions, and the result
eventually leads to the expression shown in Eq.~(\ref{eq1}).

Equation (\ref{eq1}) can be used to obtain an exact expression for the
Clebsch-Gordan coefficient as an integral.  One uses the orthogonality
of the functions $\exp(i n \theta)$ on the interval $[-\pi, \pi]$ to
select the desired coefficients in the polynomials.  Thus, we
substitute $\exp(i \theta)$ for $t$ and $\exp(i \phi)$ for $u$ in the
polynomial $(t-1)^{j+j_2-j_1} \, (t-u)^{j_1+j_2-j} \,
(u-1)^{j+j_1-j_2}$ in Eq.~(\ref{eq1}), multiply by $\exp[-i (j_1-m_1)
\phi - i (j_2-m_2) \theta]$, and integrate the two variable from
$-\pi$ to $\pi$.  The resulting expression for the Clebsch-Gordan
coefficient is
\begin{displaymath}
\langle j_1\,m_1\,j_2\,m_2 \,|\,j\,m \rangle =
(-1)^{j+m} \, N_{j_1 \, m_1 \, j_2 \, m_2 \, j \, m} \, \frac{1}{(2 \pi)^2}
\end{displaymath}
\begin{equation}
\times \int_{-\pi}^{\pi} \int_{-\pi}^{\pi}
e^{-i (j_1-m_1) \phi - i (j_2-m_2) \theta}
\, (e^{i \theta}-1)^{j+j_2-j_1} \,
(e^{i \theta}-e^{i \phi})^{j_1+j_2-j} \, (e^{i \phi}-1)^{j+j_1-j_2}
\, d\theta \, d \phi\; .
\end{equation}
This may be rewritten using the definition of the $\sin$ function,
whereupon it becomes natural to redefine the angles by a factor of
two.  The resulting form is
\begin{displaymath}
\langle j_1\,m_1\,j_2\,m_2 \,|\,j\,m \rangle =
(-1)^{j+m} \, (2 i)^{j+j_1+j_2} \, \pi^{-2} \,
N_{j_1 \, m_1 \, j_2 \, m_2 \, j \, m} \, 
\end{displaymath}
\begin{equation}
\times \int_{-\pi/2}^{\pi/2} \int_{-\pi/2}^{\pi/2}
e^{2 i m_1 \phi + 2 i m_2 \theta} \, \sin^{j+j_2-j_1} \theta \,
\sin^{j_1+j_2-j} (\theta - \phi) \, \sin^{j+j_1-j_2} \phi
\, d\theta \, d \phi\; .
\label{int}
\end{equation}
It is this integral expression for the Clebsch-Gordan coefficient
that we use in the following sections to derive formulas
for the asymptotic behavior of these coefficients.

It is also possible to express the Clebsch-Gordan coefficient as a
coefficient of a term in a polynomial in one variable, and thus as a
one-dimensional integral.  Equation~(\ref{eq1}) shows how the
Clebsch-Gordan coefficient is related to the coefficient of
$u^{j_1-m_1} \, t^{j_2-m_2}$ in the polynomial $(t-1)^{j+j_2-j_1} \,
(t-u)^{j_1+j_2-j} \, (u-1)^{j+j_1-j_2}$.  This is the same as the
coefficient of $u^{j_1-m_1} \, (u^M)^{j_2-m_2}$ in the polynomial
$(u^M-1)^{j+j_2-j_1} \, (u^M-u)^{j_1+j_2-j} \, (u-1)^{j+j_1-j_2}$
(that is, $t$ has been replaced by $u^M$) for sufficiently large
integers $M$.  This can be seen as follows.  We start by imagining the
polynomial $(t-1)^{j+j_2-j_1} \, (t-u)^{j_1+j_2-j} \,
(u-1)^{j+j_1-j_2}$ expanded out into a sum of monomials.  If $t$ is
replaced by $u^M$, each of the monomials is now just a coefficient
times a power of $u$.  We do not want any of these terms to have the
same power of $u$, otherwise they would combine and the coefficients
would change.  Thus, we look at the original polynomial
$(t-1)^{j+j_2-j_1} \, (t-u)^{j_1+j_2-j} \, (u-1)^{j+j_1-j_2}$ and ask
what the highest power of $u$ is.  This is $(j_1+j_2-j)+(j+j_1-j_2) =
2 j_1$.  We therefore select $M$ to be $2 j_1 + 1$.  The result is
that the coefficient of $u^{j_1-m_1} \, t^{j_2-m_2}$ in the polynomial
$(t-1)^{j+j_2-j_1} \, (t-u)^{j_1+j_2-j} \, (u-1)^{j+j_1-j_2}$ is the
same as the coefficient of $u^{j_1-m_1 + (2 j_1 +1)(j_2-m_2)}$ in the
polynomial $(u^{2j_1+1}-1)^{j+j_2-j_1} \, (u^{2j_1+1}-u)^{j_1+j_2-j}
\, (u-1)^{j+j_1-j_2}$.  We may drop an overall factor of
$u^{j_1+j_2-j}$, so this coefficient is the same as the coefficient of
$u^{j-m + 2 j_1(j_2-m_2)}$ in the polynomial
$(u^{2j_1+1}-1)^{j+j_2-j_1} \, (u^{2j_1}-1)^{j_1+j_2-j} \,
(u-1)^{j+j_1-j_2}$.  The resulting expression for the Clebsch-Gordan
coefficient as a coefficient in a polynomial in one variable is
\begin{displaymath}
\langle j_1\,m_1\,j_2\,m_2 \,|\,j\,m \rangle =
(-1)^{j+m} \, N_{j_1 \, m_1 \, j_2 \, m_2 \, j \, m} \,
\frac{1}{[j-m + 2 j_1(j_2-m_2)]!}
\end{displaymath}
\begin{equation}
\times
\left(\frac{d}{du}\right)^{j-m + 2 j_1(j_2-m_2)} \,
\left. \left[(u^{2j_1+1}-1)^{j+j_2-j_1} \,
(u^{2j_1}-1)^{j_1+j_2-j} \, (u-1)^{j+j_1-j_2} \right] \, \right|_{u=0} \; .
\label{eq2}
\end{equation}
As above, the selection of the coefficient in the polynomial can
also be carried out with an integral.
\begin{displaymath}
\langle j_1\,m_1\,j_2\,m_2 \,|\,j\,m \rangle =
(-1)^{j+m} \, N_{j_1 \, m_1 \, j_2 \, m_2 \, j \, m} \, \frac{1}{2 \pi}
\end{displaymath}
\begin{equation}
\times \int_{-\pi}^{\pi} e^{- i [j-m + 2 j_1(j_2-m_2)] \phi} \,
(e^{i (2j_1+1) \phi}-1)^{j+j_2-j_1} \,
(e^{i 2j_1 \phi}-1)^{j_1+j_2-j} \, (e^{i \phi}-1)^{j+j_1-j_2} \, d \phi
\; .
\end{equation}
This may be rewritten as
\begin{displaymath}
\langle j_1\,m_1\,j_2\,m_2 \,|\,j\,m \rangle =
(-1)^{j+m} \, N_{j_1 \, m_1 \, j_2 \, m_2 \, j \, m} \,
\frac{(2 i)^{j+j_1+j_2}}{2 \pi}
\end{displaymath}
\begin{equation}
\times \int_{-\pi}^{\pi}
e^{ i (2 j_1 m_2 + m) \phi} \,
\sin^{j+j_2-j_1}[(j_1+1/2) \phi] \,
\sin^{j_1+j_2-j}(j_1 \, \phi) \,
\sin^{j+j_1-j_2}(\phi / 2) \, d \phi \; ,
\end{equation}
and this may be simplified to the form
\begin{eqnarray}
\langle j_1\,m_1\,j_2\,m_2 \,|\,j\,m \rangle &=&
(-1)^{j+m} \, N_{j_1 \, m_1 \, j_2 \, m_2 \, j \, m} \,
\pi^{-1} \, 2^{j+j_1+j_2} \label{label215} \\ \nonumber
&\times& \int_{0}^{\pi}
\cos[ (2 j_1 m_2 + m) \phi + \frac{\pi}{2}(j+j_1+j_2) ] \\ \nonumber
&\times&
\sin^{j+j_2-j_1}[(j_1+1/2) \phi] \,
\sin^{j_1+j_2-j}(j_1 \, \phi) \,
\sin^{j+j_1-j_2}(\phi / 2) \, d \phi \; .
\end{eqnarray}

Although this is a one-dimensional integral (as opposed to the
two-dimensional integral presented above), it seems to be not as
useful for the study of asymptotics because of the presence of the
magnetic quantum numbers in the argument of the cosine function.

\section{Stationary-phase approximation of integral expression for
the Clebsch-Gordan coefficient}
\label{secIII}

In order to carry out a
stationary-phase approximation of the integral expression for
the Clebsch-Gordan coefficient presented in the previous section,
we begin by writing the expression
in Eq.~(\ref{int}) in the form
\begin{equation}
\langle j_1\,m_1\,j_2\,m_2 \,|\,j\,m \rangle =
(-1)^{j+m} \, (2 i)^{j+j_1+j_2} \, \pi^{-2} \,
N_{j_1 \, m_1 \, j_2 \, m_2 \, j \, m} \, 
\int_{-\pi/2}^{\pi/2} \int_{-\pi/2}^{\pi/2}
e^{g(\theta, \phi)} \, d\theta \, d \phi\; .
\label{e^g-integral}
\end{equation}
where the function $g(\theta, \phi)$ is defined to be
\begin{eqnarray}
g(\theta, \phi) &=& 2 i m_1 \phi + 2 i m_2 \theta
+ (j+j_2-j_1) \ln (\sin \theta) \\ \nonumber
&+& (j_1+j_2-j) \ln [\sin (\theta - \phi)]
+ (j+j_1-j_2) \ln (\sin \phi) \, .
\label{gdefinition}
\end{eqnarray}
Note that $g(\theta, \phi)$ has singularities
where it goes to $-\infty$,
but the integral is still well-defined because the integrand is $\exp(g)$.

To find the stationary-phase points (as explained in Appendix~A), we must first
compute the first derivatives of the function $g$.
\begin{equation}
\frac{\partial g}{\partial \theta} = 2 i m_2
+ (j+j_2-j_1) \cot \theta
+ (j_1+j_2-j) \cot (\theta - \phi) \, ,
\label{dg/dtheta}
\end{equation}

\begin{equation}
\frac{\partial g}{\partial \phi} = 2 i m_1
- (j_1+j_2-j) \cot (\theta - \phi)
+ (j+j_1-j_2) \cot \phi \, .
\end{equation}
Setting these first derivatives equal to zero results in a system of
two equations in two variables.  The identity
\begin{equation}
\cot(\theta - \phi) = \frac{1 + \cot \theta \, \cot \phi}
{\cot \phi - \cot \theta}
\label{identity19}
\end{equation}
may be used to transform this system to an
equivalent system.
\begin{equation}
2 i m + (j+j_2-j_1) \cot \theta + (j+j_1-j_2) \cot \phi = 0
\end{equation}

\begin{equation}
2 i m_2 + (j+j_2-j_1) \cot \theta
+ (j_1+j_2-j) \frac{1 + \cot \theta \, \cot \phi}{\cot \phi - \cot \theta} = 0.
\end{equation}

In order to be clear on phase conventions, choices of signs and branch cuts
we write out the steps involved in solving this system of two equations
for $\cot \theta$ and $\cot \phi$.  We start by
multiplying the second equation by $(j+j_1-j_2)(\cot \phi - \cot \theta)$
and substituting in first one:
\begin{displaymath}
[2 i m_2 + (j+j_2-j_1) \cot \theta] \{-[2 i m + (j+j_2-j_1) \cot \theta]
- (j+j_1-j_2) \cot \theta \}
\end{displaymath}
\begin{equation}
+ (j_1+j_2-j) \{(j+j_1-j_2) - \cot \theta [2 i m + (j+j_2-j_1) \cot \theta
]\}
 = 0.
\end{equation}

This is a quadratic equation in $\cot \theta$.
\begin{displaymath}
\cot^2 \theta [(j+j_2-j_1) (- 2j) - (j_1+j_2-j)(j+j_2-j_1)] +
\cot \theta [(j+j_2-j_1) (-2 i m) + 2 i m_2 (-2j)
\end{displaymath}
\begin{equation}
+ (j_1+j_2-j)(-2 i m)] + 2 i m_2 (-2 i m) + (j_1+j_2-j)(j+j_1-j_2) = 0 \, .
\end{equation}

Simplifying this results in
\begin{displaymath}
-\cot^2 \theta (j+j_2-j_1) (j_1+j_2+j)
\end{displaymath}
\begin{equation}
- 4 i \cot \theta (j_2 m + m_2 j) + 4 m_2 m + (j_1+j_2-j)(j+j_1-j_2) = 0 \, .
\end{equation}

The two solutions for the quantities $\cot \theta$ and $\cot \phi$ are
(the upper choice of sign is one solution and the lower choice of sign
is the other).
\begin{eqnarray}
\cot \theta &=& \frac{-2 i (j_2 m + m_2 j) \mp \beta}
{(j_1+j_2+j)(j+j_2-j_1)}  \, ,\nonumber \\
\cot \phi &=& \frac{-2 i (j_1 m+m_1 j) \pm \beta}
{ (j_1+j_2+j)(j+j_1-j_2)} \, ,
\label{cotangents}
\end{eqnarray}
where $\beta$ is defined to be
\begin{equation}
\beta = \sqrt{4 m_1 m_2 j^2 - 4 m m_1j_2^2 - 4 m m_2 j_1^2 + 
(j_1+j_2-j)(j+j_2-j_1)(j+j_1-j_2)(j_1+j_2+j)} \, .
\label{beta}
\end{equation}
In this equation, we use the usual choice of branch cut for the
square-root function: if the argument is negative, then the result is
a positive number times the imaginary unit.  As discussed in
Sec.~\ref{ar}, the quantity $\beta$ is real for classically allowed
sets of quantum numbers, and it is pure imaginary for classically
forbidden sets of quantum numbers.  It should be noted that this is
the same definition for the symbol $\beta$ as in Ref.~\onlinecite{BM}.

The stationary-phase approximation of the integral
$\int_{-\pi/2}^{\pi/2} \int_{-\pi/2}^{\pi/2}
e^{g(\theta, \phi)} \, d\theta \, d \phi$ that appears in the
expression for the Clebsch-Gordan coefficient in Eq.~(\ref{e^g-integral})
is given by a sum of terms of the form
\begin{equation}
\frac{2 \pi}{\sqrt{\det \frac{\partial^2 g}{\partial (\theta, \phi)^2}}} \,
e^{g(\theta, \phi)} \, ,
\label{term}
\end{equation}
summed over stationary-phase points.  The branch cut for the square
root function is just below the negative imaginary axis, as is usual.
The symbol $\frac{\partial^2 g}{\partial (\theta, \phi)^2}$ denotes
the $2\times2$ Hessian matrix of second-order derivatives of the
function $g(\theta, \phi)$, whose entries are given by
\begin{eqnarray}
\frac{\partial^2 g}{\partial \theta^2} &=&
- (j+j_2-j_1) \csc^2 \theta
- (j_1+j_2-j) \csc^2 (\theta - \phi) \, , \nonumber \\
\frac{\partial^2 g}{\partial \theta \, \partial \phi} &=&
(j_1+j_2-j) \csc^2 (\theta - \phi) \, , \nonumber \\
\frac{\partial^2 g}{\partial \phi^2} &=&
- (j_1+j_2-j) \csc^2 (\theta - \phi)
- (j+j_1-j_2) \csc^2 \phi \, .
\label{secondderivs}
\end{eqnarray}
Using the identity $\csc^2 \theta = 1 + \cot^2 \theta$ these
quantities can be expressed in terms of the cotangents in
Eq.~(\ref{cotangents}) without addressing the issue of branch cuts of
the arc-cotangent function.  The value of $\csc^2 (\theta - \phi)$ can
be determined from the quantities in Eq.~(\ref{cotangents}) using the
identity $\sin (\theta - \phi) = \sin \theta \, \cos \phi - \sin \phi
\, \cos \theta = \sin \theta \, \sin \phi (\cot \phi - \cot \theta)$.
The determinant becomes
\begin{eqnarray}
\det \frac{\partial^2 g}{\partial (\theta, \phi)^2} &=&
(j+j_2-j_1) \csc^2 \theta (j+j_1-j_2) \csc^2 \phi \nonumber \\
&+& (j_1+j_2-j) \csc^2 (\theta - \phi) [
(j+j_2-j_1) \csc^2 \theta + (j+j_1-j_2) \csc^2 \phi] \, , \nonumber \\
&=&
(1+\cot^2 \theta) (1+\cot^2 \phi) \{(j+j_2-j_1) (j+j_1-j_2) \nonumber \\
&+& \frac{(j_1+j_2-j)}{(\cot \phi - \cot \theta)^2}
[(j+j_2-j_1) (1+\cot^2 \theta) + (j+j_1-j_2) (1+\cot^2 \phi)] \} \, .
\label{det}
\end{eqnarray}
In this form the determinant is expressed entirely in term of the
cotangents of $\theta$ and $\phi$.  The quantity $e^{g(\theta, \phi)}$
can also be expressed in this way.  Choices of branch cuts are not
necessary when expressing $\sin \theta$ and $\sin \phi$ in terms of
the cotangents because only even powers of the sine-functions appear.
Using the identity
\begin{equation}
e^{i \theta} = \cos \theta + i \sin \theta = \sin \theta ( \cot \theta + i) \, ,
\end{equation}
we obtain for the factor $e^{g(\theta, \phi)}$ in Eq.~(\ref{term})
\begin{displaymath}
e^{2 i m_1 \phi + 2 i m_2 \theta} \, \sin^{j+j_2-j_1} \theta \,
\sin^{j_1+j_2-j} (\theta - \phi) \, \sin^{j+j_1-j_2} \phi
\end{displaymath}
\begin{eqnarray}
&=& (i + \cot \phi)^{2 m_1}
(i + \cot \theta)^{2 m_2}
\, \sin^{j+j_2-j_1+2 m_2} \theta \,
\sin^{j_1+j_2-j} (\theta - \phi) \, \sin^{j+j_1-j_2+2 m_1} \phi
\nonumber \\
&=& (i + \cot \phi)^{2 m_1}
(i + \cot \theta)^{2 m_2}
\, \sin^{2 j_2+2 m_2} \theta \,
(\cot\phi - \cot \theta)^{j_1+j_2-j}
\, \sin^{2 j_1+2 m_1} \phi
\nonumber \\
&=& (i + \cot \phi)^{2 m_1}
(i + \cot \theta)^{2 m_2}
\, (1 + \cot^2 \theta)^{-j_2 - m_2}
(\cot\phi - \cot \theta)^{j_1+j_2-j}
\, (1 + \cot^2 \phi)^{-j_1 - m_1}
\nonumber \\
&=& \frac{(i + \cot \phi)^{m_1-j_1}}{(-i + \cot \phi)^{j_1 + m_1}} \,
\frac{(i + \cot \theta)^{m_2-j_2}}{(-i + \cot \theta)^{j_2 + m_2}} \,
(\cot\phi - \cot \theta)^{j_1+j_2-j}
\label{e}
\end{eqnarray}
Using this equation and Eq.~(\ref{det}), all of the quantities in the
expression in Eq.~(\ref{term}) can be expressed in term of the
cotangents of $\theta$ and $\phi$, given in Eq.~(\ref{cotangents}).
It should be noted that all of the exponents in
Eq.~(\ref{e}) are integers, so choices of branch cuts are not necessary.

\subsection{Allowed region}
\label{ar}

It is useful to introduce the concepts of a triangle-allowed region
and a classically allowed region of the space of values for the
quantum numbers.  We define the triangle-allowed region to be the set
of quantum numbers for which $j_1$, $j_2$ and $j$ satisfy the triangle
inequalities and for which the inequalities $|m| \le j$ and $\{ |m_i|
\le j_i, \; i = 1,2 \}$ hold.  The Clebsch-Gordan coefficient is zero
outside of this region, so it is only within this region that
asymptotic expressions are desired.  The triangle-allowed region is
divided into a classically allowed region and a classically forbidden
region.  As is usual, we call these the allowed and forbidden regions
for brevity.  The allowed region is defined to be the set of quantum
numbers for which it is possible to define ${\bf j}$-vectors in a
three-dimensional space in such a way that their lengths are equal to
the $j$-values and their $z$-components are equal to the $m$-values
(and, of course, such that ${\bf j} = {\bf j}_1 + {\bf j}_2$).  An
example of such a construction for a set of allowed quantum numbers is
shown in Fig.~1.  It follows from the definition that the allowed
region is contained in the triangle-allowed region.  Examples of
classically forbidden points are easily found in extreme cases, such
as $m_1 = j_1$.  In this case, there is only one classically allowed
value for $m_2$ (assuming a set of triangle-allowed $j$-values have
been given), because the ${\bf j}_1$-vector must point in the
$z$-direction, and thus the $j$-triangle lies in a vertical plane.

The allowed region is the same as the region in which the three
$\lambda$-values defined in Eq.~(\ref{alpha-lambda}) satisfy the
triangle inequalities.  This is because the $\lambda$-values are the
lengths of the projections of the ${\bf j}$-vectors into the
$xy$-plane.  If the $\lambda$-values satisfy the triangle
inequalities, then it is possible to draw a triangle in the $xy$-plane
with sides equal to the $\lambda$-values.  From this, one can
construct the ${\bf j}$-vectors by simply including the $m$-values as
$z$-components.  Conversely, if the ${\bf j}$-vectors can be
constructed, then their projections into the $xy$-plane form a
triangle (with the tail of ${\bf j}_2$ at the tip of ${\bf j}_1$), and
the $\lambda$-values satisfy the triangle inequalities.

It is explained later in this paper that the $\lambda$-values satisfy
the triangle inequalities if and only if the quantity $(-\lambda_1 +
\lambda_2 + \lambda_3) (\lambda_1 - \lambda_2 + \lambda_3) (\lambda_1
+ \lambda_2 - \lambda_3)$ is nonnegative (that is, it is not possible
for two of the factors to be negative).  This observation together
with the fact that the quantity $\beta$ defined in Eq.~(\ref{beta})
may be written as
\begin{equation}
\beta =
\sqrt{(\lambda_1 + \lambda_2 + \lambda_3) (-\lambda_1 + \lambda_2 + \lambda_3)
(\lambda_1 - \lambda_2 + \lambda_3) (\lambda_1 + \lambda_2 - \lambda_3)}
\label{beta-lambda}
\end{equation}
leads us to the result that the sign of $\beta^2$ distinguishes the
allowed and forbidden regions: it is positive in the allowed region,
and it is negative in the forbidden region.
In the allowed region, $\beta$ is four times the area of the
triangle whose sides are the $\lambda$-values.  This triangle is the
projection
of the $j$-triangle into the $xy$-plane (see Fig.~1).
From Eq.~(\ref{beta}) it
is apparent that for fixed values of the $j$ quantum numbers,
$\beta^2$ is a quadratic polynomial in the $m$ quantum numbers.  Thus,
in the $(m_1,m_2)$ plane the boundary between the allowed and
forbidden regions is an ellipse.  This is shown shown in Fig.~2 for
one choice of values for $j_1$, $j_2$ and $j$.  The boundary of the
triangle-allowed region is the irregular hexagon.  The forbidden
region is composed of six subregions.  The points that separate them
are indicated in Fig.~2.  These are the points where the ellipse that
separates the allowed and forbidden regions is tangent to the hexagon
that defines the triangle-allowed region.  The coordinates of these
points can be calculated from the expression for $\beta$ and the
equations for the straight-line sections of the boundary of the
triangle-allowed region.  The resulting coordinates of these points
are indicated in the figure.

The calculations involved in the stationary-phase approximation of the
integral expression for the Clebsch-Gordan coefficient are different
in the allowed and forbidden regions.  We will treat the allowed
region first.  The sum over stationary-phase points for the case where
the set of quantum numbers is in the allowed region is a sum over both
of the two solutions for the cotangents of $\theta$ and $\phi$ given
in Eq.~(\ref{cotangents}).  This is analogous to the behavior
demonstrated in Appendix~A, and it is also the same as in the
calculation of the stationary-phase approximation of the Airy
integral, which is the canonical example of a stationary-phase
calculation.  In the case of the Airy integral, there are allowed and
forbidden regions in position space, and in the allowed region the
contour of integration is deformed to run over both of the
stationary-phase points.  We define $c_\theta$ and $c_\phi$ to be the
first solution for the cotangents in Eq.~(\ref{cotangents}).
\begin{eqnarray}
c_\theta &=& \frac{-2 i (j_2 m + m_2 j) - \beta}
{(j+j_2-j_1) (j_1+j_2+j)}  \, ,\nonumber \\
c_\phi &=& \frac{-2 i (j_1 m+m_1 j) + \beta}
{ (j_1+j_2+j)(j+j_1-j_2)} \, ,
\label{cs}
\end{eqnarray}
Because of the form of the two solutions given in
Eq.~(\ref{cotangents}), the second solution is obtained from this one
by multiplying by minus one and complex conjugating.  Note this is
only valid in the allowed region, where the quantity $\beta$ is real.
The first term in the sum over stationary-phase points is given by
plugging the expression for $\det \frac{\partial^2 g}{\partial
(\theta, \phi)^2}$, given in Eq.~(\ref{det}), and the expression for
$e^{g(\theta, \phi)}$, given in Eq.~(\ref{e}), into the quantity in
Eq.~(\ref{term}), using $c_\theta$ and $c_\phi$ for the cotangents.
The second term in the sum over stationary-phase points is the same,
except $-c_\theta^\ast$ and $-c_\phi^\ast$ are used for the
cotangents.  The result for the determinant in the second term is
obtained by simply complex conjugating the first value, since all of
the cotangents in this expression are squared.  As for the
$e^{g(\theta, \phi)}$ factor in the second term, we start by
considering the expression for this factor in the first term:
\begin{equation}
\frac{(i + c_\phi)^{m_1-j_1}}{(-i + c_\phi)^{j_1 + m_1}} \,
\frac{(i + c_\theta)^{m_2-j_2}}{(-i + c_\theta)^{j_2 + m_2}} \,
(c_\phi - c_\theta)^{j_1+j_2-j} \, .
\end{equation}
The complex conjugate of this is
\begin{displaymath}
\frac{(-i + c_\phi^\ast)^{m_1-j_1}}{(i + c_\phi^\ast)^{j_1 + m_1}} \,
\frac{(-i + c_\theta^\ast)^{m_2-j_2}}{(i + c_\theta^\ast)^{j_2 + m_2}} \,
(c_\phi^\ast - c_\theta^\ast)^{j_1+j_2-j}
\end{displaymath}
\begin{equation}
= (-1)^{j_1+j_2+j} \,
\frac{(i - c_\phi^\ast)^{m_1-j_1}}{(-i - c_\phi^\ast)^{j_1 + m_1}} \,
\frac{(i - c_\theta^\ast)^{m_2-j_2}}{(-i - c_\theta^\ast)^{j_2 + m_2}} \,
(-c_\phi^\ast + c_\theta^\ast)^{j_1+j_2-j}
\label{20}
\end{equation}
This shows that the $e^{g(\theta, \phi)}$ factor in the second term
[which appears after the $(-1)^{j_1+j_2+j}$ in the last line of
Eq.~(\ref{20})] is $(-1)^{j_1+j_2+j}$ times the complex conjugate of
the $e^{g(\theta, \phi)}$ factor in the first term.  Thus the second
term is $(-1)^{j_1+j_2+j}$ times the complex conjugate of the first
term.  It is therefore convenient to obtain the sum over
stationary-phase points by taking $i^{j_1+j_2+j}$ times the first
term, adding the complex conjugate of this product, and then dividing
by $i^{j_1+j_2+j}$.  Using the fact that the real part of a quantity
$x$ is given by $\Re[x] = (x + x^\ast)/2$, our stationary-phase
approximation for the integral expression for the Clebsch-Gordan
coefficient in Eq.~(\ref{e^g-integral}) can be written as
\begin{eqnarray}
\langle j_1\,m_1\,j_2\,m_2 \,|\,j\,m \rangle &\approx &
(-1)^{j+m} \, (2 i)^{j+j_1+j_2} \, \pi^{-2} \,
N_{j_1 \, m_1 \, j_2 \, m_2 \, j \, m} 
\frac{2}{i^{j_1+j_2+j}} \Re \left [ 
\frac{2 \pi \, i^{j_1+j_2+j}}
{\sqrt{\det \frac{\partial^2 g}{\partial (\theta, \phi)^2}}} \,
e^{g(\theta, \phi)} \right ] \nonumber \\
&= & (-1)^{j+m} \, 2^{j+j_1+j_2+2} \, \pi^{-1} \,
N_{j_1 \, m_1 \, j_2 \, m_2 \, j \, m} \, 
\Re \left [ \frac{i^{j_1+j_2+j}}
{\sqrt{\det \frac{\partial^2 g}{\partial (\theta, \phi)^2}}} \,
e^{g(\theta, \phi)} \right ] \, ,
\label{CGpaperallowed}
\end{eqnarray}
where the quantities $\det \frac{\partial^2 g}{\partial (\theta, \phi)^2}$
and $e^{g(\theta, \phi)}$ are obtained from Eq.~(\ref{det}) and
Eq.~(\ref{e}) using the $c_\theta$ and $c_\phi$ given in Eq.~(\ref{cs}).

Although the expression in Eq.~(\ref{CGpaperallowed}) gives a value
that is a real number, it involves intermediate quantities that are
complex.  It is possible to transform this expression so that only
real quantities are involved.  This transformation is very lengthy,
and it is not practical to describe it in detail here.  Instead, we
present an expression that is exactly equal to the expression in
Eq.~(\ref{CGpaperallowed}) in the allowed region.  This equality can
be verified most convincingly by substituting numerical values into
the expressions and evaluating the results to high numerical precision
(much higher than the level at which discrepancies would occur if
order $\hbar$ terms were dropped).  A brief description of the
transformation is the following.  Every complex quantity $x+iy$ that
occurs in Eq.~(\ref{CGpaperallowed}) is written as the product of a
modulus and a phase, $\sqrt{x^2+y^2} \, \exp[i \tan^{-1}(y/x)]$, where
care must be taken that correct branches are used for each $x+iy$,
that is, one must examine the quantities $x$ and $y$ to determine the
range of phase factors $(x+iy)/\sqrt{x^2+y^2}$ that can occur in the
allowed region, and make branch choices accordingly.  At some stages
in the calculation, large polynomials are involved, and computer-aided
symbol manipulation becomes useful in working with these.  Our result
may be put in the form
\begin{equation}
\langle j_1\,m_1\,j_2\,m_2 \,|\,j\,m \rangle \approx
2 I_{j_1 \, m_1 \, j_2 \, m_2 \, j \, m} \,
\sqrt{\frac{j}{\pi \beta}} \, \cos \left[\chi + \frac{\pi}{4} - \pi(j+1) \right]
\, ,
\label{ourexpr}
\end{equation}
where $\chi$ is defined to be
\begin{eqnarray}
\chi &=&
\left(j_1+\frac{1}{2} \right)
\cos^{-1}\left[{\frac{(-m)(j_1^2 + j_2^2 - j^2) - m_2(j_1^2 + j^2 -j_2^2)}
{\alpha \lambda_1}}\right] \nonumber \\
&+&
\left(j_2+\frac{1}{2} \right)
\cos^{-1}\left[{\frac{m_1(j^2 + j_2^2 - j_1^2) - (-m)(j_2^2 + j_1^2 -j^2)}
{\alpha \lambda_2}}\right] \nonumber \\
&+&
\left(j+\frac{1}{2} \right)
\cos^{-1}\left[{\frac{m_2(j_1^2 + j^2 - j_2^2) - m_1(j^2 + j_2^2 -j_1^2)}
{\alpha \lambda_3}}\right] \nonumber \\
&-&
m_1 \cos^{-1}\left[{\frac{\lambda_1^2 + \lambda_3^2 - \lambda_2^2}
{2 \lambda_1 \lambda_3}}\right] \nonumber \\
&+&
m_2 \cos^{-1}\left[{\frac{\lambda_3^2 + \lambda_2^2 - \lambda_1^2}
{2 \lambda_2 \lambda_3}}\right] \, ,
\label{chi}
\end{eqnarray}
and
\begin{eqnarray}
\alpha &=&
\sqrt{(j + j_1 + j_2) (-j + j_1 + j_2) (j - j_1 + j_2) (j + j_1 - j_2)}
\nonumber \\
\lambda_i &=& \sqrt{j_i^2-m_i^2} \; \; \; \; \; i = 1,2 \nonumber \\
\lambda_3 &=& \sqrt{j^2-m^2} \, .
\label{alpha-lambda}
\end{eqnarray}
(The quantity $\alpha$ is four times the area of the $j$-triangle
shown in Fig.~1.)
Note that the $\cos^{-1}$ functions in Eq.~(\ref{chi}) are the usual
principal branch, whose range is the interval from zero to $\pi$.
The quantity $I_{j_1 \, m_1 \, j_2 \, m_2 \, j \, m}$ is defined to be
\begin{eqnarray}
I_{j_1 \, m_1 \, j_2 \, m_2 \, j \, m} &=&
\sqrt{ \frac{(j+1/2)(j+j_1+j_2)}{j\,(j+j_1+j_2+1)} } \nonumber \\
&\times&
\frac{f(j_1+m_1)\, f(j_1-m_1)\, f(j_2+m_2)\, f(j_2-m_2)\,
f(j+m)\,f(j-m)}
{f(j_1+j_2+j)\, f(j_1+j_2-j)\, f(j_1-j_2+j)\, f(-j_1+j_2+j)} \, ,
\end{eqnarray}
where the function $f$ is defined to be
\begin{equation}
f(n) = \sqrt{ \frac{n!}{\sqrt{2 \pi n} \, n^n \, e^{-n}} } \, ,
\end{equation}
that is, $f(n)$ is the square root of the ratio of $n!$ to the
Stirling approximation of $n!$.  Note that for large $n$, $f(n)$
approaches one.  Thus, for large quantum numbers, $I_{j_1 \, m_1 \,
j_2 \, m_2 \, j \, m}$ approaches one.  It differs from unity by a
correction that is order $\hbar$, as can be deduced from the
discussion of the Stirling approximation in Appendix~A.  As mentioned
above, we present our approximation in the form given in
Eq.~(\ref{ourexpr}) so that the exact equality of this expression and
the complex expression given in Eq.~(\ref{CGpaperallowed}) can be
verified numerically.  The factor $I_{j_1 \, m_1 \, j_2 \, m_2 \, j \,
m}$ may be dropped without reducing the quality of the approximation,
that is, the ratio of our approximation to the exact value differs
from unity by a quantity that is order $\hbar$.  Thus, we may write
our approximation in the form
\begin{equation}
\langle j_1\,m_1\,j_2\,m_2 \,|\,j\,m \rangle \approx
2  \, \sqrt{\frac{j}{\pi \beta}} \, \cos \left[\chi - \pi \left(j+\frac{3}{4}\right) \right] \, .
\label{ourexprsimp}
\end{equation}
Ponzano and Regge\cite{PR} give a geometrical interpretation of the
five angles that occur in the expression for $\chi$ in
Eq.~(\ref{chi}).
An equation similar to Eq.~(\ref{ourexprsimp}) also appears in
Ref.~\onlinecite{BM}, but the $(j+1/2)$ factors in $\chi$ are
included at the end of the calculation to improve the accuracy, and
the $\pi(j+1)$ in
Eq.~(\ref{ourexpr}) is missing so that the formula gives the wrong sign
for even $j$-values and does not give
the right magnitude for half-integer $j$ values.

\subsection{Forbidden region}

In the forbidden region, only one of the stationary-phase points is
used in the approximation.  This is analogous to the situation in the
Airy-function problem mentioned above, where there are dominant and
subdominant branches, and in the forbidden region only the subdominant
branch exists.  Similarly, the model problem in Appendix~A shows how
for the case of $m > n$, two stationary-phase points are used, while
for the case $m < n$ only one stationary-phase point is involved.  The
choice of which of the two roots in Eq.~(\ref{cotangents}) is to be
used for our approximation of the Clebsch-Gordan coefficient is
indicated in Table~I.  Given the $m$-values of a point in the
forbidden region in Fig.~2, it is inconvenient to determine which
subregion it is in by using nested if-then statements, because the
relative ordering of, say, the $m_2$-coordinates of the points on the
boundaries between the forbidden subregions changes as the $j$-values
are changed.  A much simpler way to determine which branch to use is
to find the sign of a certain polynomial which we describe here.  As
can be seen from Table~I, the choice of branch alternates as one goes
around the diagram in Fig.~2.  Thus we use the sign of the product of
three expressions that flip signs in the right way.  Given the
coordinates of one of the boundary points in the $(m_1, m_2)$-plane, a
vector perpendicular to it can be constructed by exchanging the
coordinates and changing the sign of one of them.  The dot-product of
this vector and $(m_1, m_2)$ is a function on the $(m_1, m_2)$-plane
that changes sign at the boundary between the two subregions in
question.  Thus we are led to consider the sign of the function
\begin{displaymath}
\left[(m_1,m_2)  \cdot  (-2 j_2^2, j^2 - j_1^2 - j_2^2) \right]
\end{displaymath}
\begin{displaymath}
\times \left[(m_1,m_2)  \cdot  (-j^2+j_1^2-j_2^2, j^2+j_1^2-j_2^2) \right] 
\end{displaymath}
\begin{displaymath}
\times \left[(m_1,m_2)  \cdot  (-j^2+j_1^2+j_2^2, 2 j_1^2) \right] \, .
\end{displaymath}
If this quantity is positive (negative), then the upper (lower) choice
of root in Eq.~(\ref{cotangents}) is used.  Once the cotangents of the
angles at the stationary-phase point are determined, the approximation
of the Clebsch-Gordan coefficient can be evaluated from the expression
\begin{equation}
\langle j_1\,m_1\,j_2\,m_2 \,|\,j\,m \rangle \approx 
(-1)^{j+m} \, (2 i)^{j+j_1+j_2} \, \pi^{-2} \,
N_{j_1 \, m_1 \, j_2 \, m_2 \, j \, m}
\frac{2 \pi \, e^{g(\theta, \phi)}}
{\sqrt{\det \frac{\partial^2 g}{\partial (\theta, \phi)^2}}} \, .
\label{complex-forbidden}
\end{equation}
All of the quantities needed to evaluate this expression were
expressed in terms of the cotangents of the angles in Eqs.~(\ref{det})
and (\ref{e}).  It may be noted that in the forbidden region the
cotangents become pure imaginary, as can be seen from
Eq.~(\ref{cotangents}).  This behavior is similar to that in the model
problem in Appendix~A, where the angle suddenly jumps in terms of its
real part (but the analogy is not perfect because in the model problem
the cotangent is pure imaginary in both the region $m>n$ and the
region $m<n$).

Evaluating Eq.~(\ref{complex-forbidden}) results in a real value,
although complex numbers are involved at intermediate steps.  As in
the case of our analysis in the allowed region, the expression may be
transformed to a form that involves only operations with real numbers.
This can be done in a way that parallels the previous calculation,
with hyperbolic functions playing the role of trigonometric functions.
The transformation involves choices of branch cuts and depends on
which of the six subregions of the forbidden region one is working in.
Thus there are six different all-real expressions for the forbidden
region.  In the interest of brevity, we will present only one of these
here.  In subregion $VI$, the expression in
Eq.~(\ref{complex-forbidden}) is exactly equal to
\begin{equation}
(-1)^{j_2+m_2} \, 2 \, I_{j_1 \, m_1 \, j_2 \, m_2 \, j \, m} \,
\sqrt{\frac{j}{\pi \, |\beta|}} \, \exp(- \chi^{(vi)}) \, ,
\label{ourexpr2}
\end{equation}
where $\chi^{(vi)}$ is defined to be
\begin{eqnarray}
\chi^{(vi)} &=&
\left(j_1+\frac{1}{2} \right)
\cosh^{-1}\left[{\frac{-m(j_1^2 + j_2^2 - j^2) - m_2(j_1^2 + j^2 -j_2^2)}
{\alpha \lambda_1}}\right] \nonumber \\
&-&
\left(j_2+\frac{1}{2} \right)
\cosh^{-1}\left[{\frac{-m_1(j^2 + j_2^2 - j_1^2) - m(j_2^2 + j_1^2 -j^2)}
{\alpha \lambda_2}}\right] \nonumber \\
&-&
\left(j+\frac{1}{2} \right)
\cosh^{-1}\left[{\frac{-m_2(j_1^2 + j^2 - j_2^2) + m_1(j^2 + j_2^2 -j_1^2)}
{\alpha \lambda_3}}\right] \nonumber \\
&-&
m \cosh^{-1}\left[{\frac{\lambda_1^2 + \lambda_3^2 - \lambda_2^2}
{2 \lambda_1 \lambda_3}}\right] \nonumber \\
&-&
m_2 \cosh^{-1}\left[{\frac{\lambda_1^2 + \lambda_2^2 - \lambda_3^2}
{2 \lambda_2 \lambda_1}}\right] \, ,
\label{chi2}
\end{eqnarray}
This all-real expression was derived by a very lengthy calculation, as
in the case of the analysis in the allowed region.  Again, an exact
equality such as the one above can be checked easily by substituting
in test numbers and evaluating to sufficient precision.  As before, to
actually use the approximation, one would drop the factor of $I_{j_1
\, m_1 \, j_2 \, m_2 \, j \, m}$ since it can be approximated by
unity, to the order that we are working in this section.  All-real
expressions for the other subregions of the forbidden region can most
easily be obtained by using the symmetries of the Clebsch-Gordan
coefficients to related the expressions for the different subregions.
If one prefers not to work with six different expressions for the
forbidden region, one can use the polynomial discussed above to select
the required stationary-phase point and then plug this into the
approximation given in Eq.~(\ref{complex-forbidden}).  This requires
operations with complex numbers, but is easier to implement in a
computer program.  Alternatively, to obtain an approximate value for
the Clebsch-Gordan coefficient for a given point in the forbidden
region, one could work with only one all-real expression for a
particular forbidden subregion and use the symmetries of the
Clebsch-Gordan coefficients to map the given point to a point that is
within the subregion for which the expression is valid.

It is interesting to compare the all-real expressions obtained in the
allowed region, Eq.~(\ref{ourexpr}), and in the forbidden region,
Eq.~(\ref{ourexpr2}).  They are similar in form, but the behavior is
oscillatory in the allowed region and exponentially decaying in the
forbidden region.  This is the behavior expected in quantum mechanical
problems that have an allowed region and a forbidden region.

In the forbidden region, writing the approximation in an all-real form
is illuminating because it makes it apparent that sign functions
exist.  We call the factor $(-1)^{j_2+m_2}$ in Eq.~(\ref{ourexpr2}) a
sign function.  The remaining factors in that equation are all
positive, so the sign function gives the sign of the result.  However,
since the result is an approximation of a Clebsch-Gordan coefficient,
the sign function also gives the sign of the Clebsch-Gordan
coefficient, at least in the asymptotic regime.  Thus, the sign
functions are actually properties of the Clebsch-Gordan coefficients
themselves, for a given choice of phase conventions.  We are using the
conventions defined by Eq.~(\ref{Wigner}).  The existence of sign
functions was not clear from Eq.~(\ref{Wigner}), which was our
starting point.  The sign functions for each of the six forbidden
subregions are given in Table~I.

The angle $\chi$, given in Eq.~(\ref{chi}), that appears in our
approximation in the allowed region can be rewritten in several
different ways.  The reason is that the angles that multiply the $m$'s
in the equation for $\chi$ are two of the interior angles in the
triangle formed by the three $\lambda$-values.  If we call these
angles $\alpha_1$, $\alpha_2$ and $\alpha_3$ (where $\alpha_i$ is the
angle opposite the side of length $\lambda_i$), then we have
\begin{eqnarray}
m_1 + m_2 &=& m \, , \nonumber \\
\alpha_1 + \alpha_2 + \alpha_3 &=& \pi \, .
\end{eqnarray}
Thus, the vectors $(m_1,m_2,-m)$ and
$(\alpha_1,\alpha_2,\alpha_3-\pi)$ are both perpendicular to
$(1,1,1)$, and their cross-product is parallel to $(1,1,1)$.  Each of
the three components of their cross product are thus equal, and each
one could be used as part of $\chi$ in the allowed region,
\begin{equation}
m_1 \alpha_2 - m_2 \alpha_1 = m_2 (\alpha_3-\pi) + m \alpha_2 = -m \alpha_1 - m_1 (\alpha_3-\pi) \, .
\end{equation}

In the forbidden region there is no such flexibility in how to write
the corresponding terms, because there do not exist three angles
corresponding to $\alpha_1$, $\alpha_2$ and $\alpha_3$.  This is
because it is not possible to form a triangle using the three
$\lambda$-values.  Three different quantities like
\begin{equation}
\frac{\lambda_3^2+\lambda_1^2-\lambda_2^2}{2 \lambda_3 \lambda_1}
\label{perm}
\end{equation}
can be written down by cyclically permuting the indices, but only two
of these can be used as arguments of the $\cosh^{-1}$ function in an
all-real expression.  This can be seen in the following way.  The
$\lambda$'s are nonnegative and if three nonnegative numbers fail to
satisfy the triangle inequalities, exactly one triangle inequality is
violated.  [Proof: Let $\lambda_{max}$ be the largest value,
$\lambda_{mid}$ be the middle value, and $\lambda_{min}$ be the
smallest.  Then $-\lambda_{min} + \lambda_{mid} + \lambda_{max} \ge 0$
and $\lambda_{min} - \lambda_{mid} + \lambda_{max} \ge 0$, so we must
have $\lambda_{min} + \lambda_{mid} - \lambda_{max} < 0$.]  Now we
consider rewriting the expression
\begin{equation}
\frac{\lambda_3^2+\lambda_1^2-\lambda_2^2}{2 \lambda_3 \lambda_1}
=1 + \frac{(\lambda_3-\lambda_1)^2 - \lambda_2^2}
{2 \lambda_3 \lambda_1}
= 1 - \frac{(\lambda_1 + \lambda_2 - \lambda_3)(\lambda_2+\lambda_2-\lambda_1)}
{2 \lambda_3 \lambda_1} \, .
\end{equation}
This shows that of the three permutations of the expression in
Eq.~(\ref{perm}), exactly two will be greater than unity.  It is these
two that must be used as arguments of the $\cosh^{-1}$ function in an
all-real expression.  Thus there is no flexibility in ways to write
the $m$-terms in $\chi$ as in the allowed region.  Throughout each one
of the six subregions of the forbidden region, a single triangle
inequality for the $\lambda$'s is violated.  It is not possible that
one triangle inequality is violated in one part of a subregion and
another triangle inequality is violated in another part of the same
subregion because at the boundary between these two parts $\beta$
would be zero, as can be seen from Eq.~(\ref{beta-lambda}).  But
$\beta^2$ is a quadratic polynomial in $m_1$ and $m_2$ [see
Eq.~(\ref{beta})], the zero-contour of which is the ellipse in Fig.~2,
so it is not possible for it to be zero along a curve in the forbidden
region.  The $\lambda$ that is largest in each subregion is indicated
in Table~I.  The forms of all-real expressions in each of the
forbidden subregions will reflect the fact that in each one of the
subregions one of the $\lambda$'s is larger than the sum of the other
two.

It remains to discuss the case of points that are on the boundary
between the allowed and forbidden regions.  The quantity $\beta$ in
Eq.~(\ref{beta}) is zero on this boundary, and since $\beta$ is
invariant under the full 72-element symmetry group of the 3-j
symbol\cite{Regge}, the Clebsch-Gordan coefficient cannot be
approximated on the boundary with the formulas presented in this
paper.  The reason is that $\beta$ occurs in the denominator in
Eqs.~(\ref {ourexprsimp}) and (\ref{ourexpr2}).  Since $\beta^2$ is a
homogeneous polynomial in the quantum numbers, it will be zero for
sets of quantum numbers equal to any multiple of a set of quantum
numbers for which $\beta$ is zero.  The behavior of the Clebsch-Gordan
coefficients in the direction transverse to the boundary should be
similar to that of the Airy function (see Ref.~\onlinecite{PR}).

The invariance of $\beta$ under the 72-element symmetry group of the
3-j symbols may be shown as follows.  We begin by constructing the
$3\times3$ Regge array of linear combinations of quantum numbers,
given in Ref.~\onlinecite{Regge}.  For any integer $n$, we define the
polynomial $p_n$ to be the sum of the $n$-th powers of the nine
elements of this matrix.  These polynomials are invariant under the
symmetry group, because if two sets of quantum numbers are related by
a Regge symmetry, we can construct the $3\times3$ Regge array for each
set and compute $p_n$.  The results are the same because of the
commutativity of addition.  It is possible to write $\beta^2$ in terms
of the $p_n$.
\begin{equation}
\beta^2 = (p_1^4 - 6 p_2 \, p_1^2 - 27 p_2^2 + 108 p_4)/324 \,.
\end{equation}
The coefficients in this equation may be simplified slightly by using
the relation $p_1 = 3(j_1+j_2+j)$.  This equation proves the invariance
of $\beta$ under the symmetry group.

An example of quantum numbers for which $\beta$ is zero is
\begin{equation}
(j_1, m_1, j_2, m_2, j,m) = (3, -2, 6, 4, 7, 2).
\end{equation}
This point is not on the edge of the triangle-allowed region.  Points
for which $\beta$ is zero and which are on the edge of the
triangle-allowed region are easier to find.  For example, one can
choose $(j_1, m_1) = (j_2, m_2) = \frac{1}{2} (j,m)$.  For such a
point, $j_1+j_2-j$ is zero.

The reason we are unable to approximate the Clebsch-Gordan coefficient
for cases in which $\beta$ is zero is that the determinant of the $2
\times 2$ matrix of second derivatives of $g(\theta, \phi)$ is zero at
the stationary-phase points.  This can be shown by plugging the
solutions for the cotangents of $\theta$ and $\phi$ at a
stationary-phase point into Eq.~(\ref{det}) for the determinant; the
result has an overall factor of $\beta$ after being simplified [see
Eq.~(\ref{detbeta})].  This determinant appears in the denominator of
Eq.~(\ref{term}), so our method cannot be applied.  Note that when
$\beta$ is zero, the two solutions for the cotangents at the
stationary-phase point are the same [see Eq.~(\ref{cotangents})].
Also, it should be noted that if any of the $m$-values has its
absolute value close to the corresponding $j$, then the corresponding
$\lambda$ will be small [see Eq.~(\ref{alpha-lambda})], and the area
of the $\lambda$-triangle will be small.  Thus, $\beta$ will be small,
and the set of quantum numbers is close to the boundary.  In contrast
to this, there is no difficulty with the approximation if the $m$
values are close to zero.  These considerations are mirrored in the
approximation (using Stirling's formula) of $N_{j_1 \, m_1 \, j_2 \,
m_2 \, j \, m}$, defined in Eq.~(\ref{Ndefinition}); no factorials of
$m$-values appear, only factorials of $j-m$, $j+m$, etc.

\section{Higher-Order Approximation}

The methods used in the previous sections can be extended to higher
order.  In this section, we derive the next correction to the previous
results.  The approximation that is obtained in this way gives results
that are accurate to six digits, for example, when the quantum numbers
are in the hundreds.

Let $(\theta_0, \phi_0)$ be a stationary-phase point, {\it i.e.} a point
at which $\frac{\partial g}{\partial \theta} = 
\frac{\partial g}{\partial \phi} = 0$.
We write the Taylor expansion of the function $g(\theta, \phi)$
about the point $(\theta_0, \phi_0)$ as a sum of homogeneous polynomials,
\begin{equation}
g(\theta_0 + x, \phi_0 + y) = g_0 + g_2 + g_3 + g_4 + \ldots ,
\label{g4}
\end{equation}
where
\begin{eqnarray}
g_0 &=& g(\theta_0, \phi_0) \, , \nonumber \\
g_2 &=& g_{_{\theta\theta}} x^2/2 + g_{_{\theta\phi}} x y
+ g_{_{\phi\phi}} y^2/2 \, ,
\nonumber \\
g_3 &=& g_{_{\theta\theta\theta}} x^3/6 + g_{_{\theta\theta\phi}} x^2 y/2
+ g_{_{\theta\phi\phi}} x y^2/2 + g_{_{\phi\phi\phi}} y^3/6 \, , \nonumber \\
g_4 &=& g_{_{\theta\theta\theta\theta}} x^4/24
 + g_{_{\theta\theta\theta\phi}} x^3 y/6
 + g_{_{\theta\theta\phi\phi}} x^2 y^2/4 
 + g_{_{\theta\phi\phi\phi}} x y^3/6
 + g_{_{\phi\phi\phi\phi}} y^4/24 \, ,
\end{eqnarray}
where, for example, $g_{_{\theta\theta\phi}}$ is defined to be
$\partial^3 g/\partial \theta^2 \partial \phi$ at the stationary-phase point.

To obtain the next higher stationary-phase approximation for the
Clebsch-Gordan coefficient, we terminate the series in Eq.~(\ref{g4})
at the fourth-order term.  The reason for this is explained below.
Thus, the approximation of the function $g$ has derivatives at the
stationary-phase point $(\theta_0, \phi_0)$ that agree with those of
$g$ through fourth order.  Our approximation of the integrand
$\exp(g)$ is
\begin{eqnarray}
\exp[g(\theta_0 + x, \phi_0 + y)] &\approx& \exp(g_0) \exp(g_2)
\exp(g_3) \exp(g_4) \nonumber \\
&=& \exp(g_0) \exp(g_2)
(1+g_3+g_3^2/2+\ldots) (1+g_4+g_4^2/2+\ldots) \, .
\end{eqnarray}
When this is multiplied out, each of the terms may be integrated over
the $xy$-plane in closed form.  We are interested in the asymptotic
behavior of the resulting terms.  The question is how the terms behave
when all of the quantum numbers $(j_1, m_1, j_2, m_2, j, m)$ are
multiplied by the same factor (such a factor is called $1/\hbar$, as
explained in the introduction).  The stationary-phase point
$(\theta_0, \phi_0)$ is independent of the factor, {\it i.e.}
$(\theta_0, \phi_0)$ is order $\hbar^0$, as can be seen from
Eq.~(\ref{cotangents}).  The second derivatives of $g$ at $(\theta_0,
\phi_0)$ are order $1/\hbar$, as can be seen from
Eq.~(\ref{secondderivs}), so to see how the integral of $\exp(g_2)$
depends on $\hbar$, we define new variables of integration to be the
old variables times $\hbar^{-1/2}$.  From this it follows that the
integral of $\exp(g_2)$ is order $\hbar$.  By the same reasoning, the
integral of a homogeneous quartic polynomial times $\exp(g_2)$ is
order $\hbar^3$, and the integral of a homogeneous sixth-order
polynomial times $\exp(g_2)$ is order $\hbar^4$.  The polynomials
$g_4$ and $g_3^2$ have coefficients that are order $1/\hbar$ and
$1/\hbar^2$, respectively, so the integrals of these times $\exp(g_2)$
are both order $\hbar^2$.  This is one order of $\hbar$ smaller than
the integral of $\exp(g_2)$.  The integral of a homogeneous polynomial
of odd degree times $\exp(g_2)$ vanishes due to antisymmetry.  Thus,
the next higher order approximation of the integral of $\exp(g)$ is
obtained by integrating
\begin{equation}
\exp(g_0) \exp(g_2) (1+g_4+g_3^2/2) \, .
\label{g4g32}
\end{equation}
Terms coming from $g_5$, etc, contribute at higher orders.

To find the ratio of the integral of $g_4 \exp(g_2)$
to the integral of $\exp(g_2)$ the following integrals are necessary.
\begin{eqnarray}
i_1 &=& \int_{-\infty}^{\infty} \int_{-\infty}^{\infty}
\exp(g_2) \, dx \, dy \, , \nonumber \\
i_2 &=& \int_{-\infty}^{\infty} \int_{-\infty}^{\infty}
x^4 \exp(g_2) \, dx \, dy \, , \nonumber \\
i_3 &=& \int_{-\infty}^{\infty} \int_{-\infty}^{\infty}
x^3 y\exp(g_2) \, dx \, dy \, , \nonumber \\
i_4 &=& \int_{-\infty}^{\infty} \int_{-\infty}^{\infty}
x^2 y^2\exp(g_2) \, dx \, dy \, .
\end{eqnarray}

We will need the ratios
$i_2/i_1$, $i_3/i_1$ and $i_4/i_1$.  The integrals are tabulated
and these ratios can be worked out without the use of any information
about relationships between the various derivatives of the function $g$
at the stationary-phase point.  The results are
\begin{eqnarray}
i_2/i_1 &=& \frac{3 g_{_{\phi\phi}}^2}
{(g_{_{\theta\theta}} \, g_{_{\phi\phi}} - g_{_{\theta\phi}}^2)^2} \, ,
\nonumber \\
i_3/i_1 &=& \frac{-3 g_{_{\theta\phi}} \, g_{_{\phi\phi}}}
{(g_{_{\theta\theta}} \, g_{_{\phi\phi}} - g_{_{\theta\phi}}^2)^2} \, ,
\nonumber \\
i_4/i_1 &=& \frac{2 g_{_{\theta\phi}}^2 + g_{_{\theta\theta}} \,
g_{_{\phi\phi}}}
{(g_{_{\theta\theta}} \, g_{_{\phi\phi}} - g_{_{\theta\phi}}^2)^2} \, .
\end{eqnarray}
It may be noted that $g_{_{\theta\theta}} \, g_{_{\phi\phi}} -
g_{_{\theta\phi}}^2$ is the determinant of the $2\times2$ matrix of
second partial derivatives of the function $g$.

The ratio, which we denote by $\delta_4$,
of the integral of $g_4 \exp(g_2)$ to the integral of $\exp(g_2)$
works out to be
\begin{equation}
\delta_4 =
\frac{
g_{_{\theta\theta\theta\theta}} \, g_{_{\phi\phi}}^2 -
4 g_{_{\theta\theta\theta\phi}} \, g_{_{\theta\phi}} \, g_{_{\phi\phi}} +
2 g_{_{\theta\theta\phi\phi}} (
2 g_{_{\theta\phi}}^2 + g_{_{\theta\theta}} \, g_{_{\phi\phi}}) -
4 g_{_{\theta\phi\phi\phi}} \, g_{_{\theta\theta}} \, g_{_{\theta\phi}} +
g_{_{\phi\phi\phi\phi}} \, g_{_{\theta\theta}}^2
}{8 (g_{_{\theta\theta}} \, g_{_{\phi\phi}} - g_{_{\theta\phi}}^2)^2}
\label{alpha4}
\end{equation}

Next, we move on to the $g_3^2$ term in Eq.~(\ref{g4g32}).
To find the ratio of the integral of $\frac{1}{2} g_3^2 \exp(g_2)$
to the integral of $\exp(g_2)$ the following integrals are necessary.
\begin{eqnarray}
i_5 &=& \int_{-\infty}^{\infty} \int_{-\infty}^{\infty}
x^6 \exp(g_2) \, dx \, dy \, , \nonumber \\
i_6 &=& \int_{-\infty}^{\infty} \int_{-\infty}^{\infty}
x^5 y\exp(g_2) \, dx \, dy \, , \nonumber \\
i_7 &=& \int_{-\infty}^{\infty} \int_{-\infty}^{\infty}
x^4 y^2\exp(g_2) \, dx \, dy \, , \nonumber \\
i_8 &=& \int_{-\infty}^{\infty} \int_{-\infty}^{\infty}
x^3 y^3\exp(g_2) \, dx \, dy \, .
\end{eqnarray}

We will need the ratios
$i_5/i_1$, $i_6/i_1$, $i_7/i_1$ and $i_8/i_1$.
As in the case of the $g_4$ calculation, the integrals are tabulated
and these ratios can be worked out without the use of any information
about relationships between the various derivatives of the function $g$
at the stationary-phase point.  The results are
\begin{eqnarray}
i_5/i_1 &=& \frac{-15 g_{_{\phi\phi}}^3}
{(g_{_{\theta\theta}} \, g_{_{\phi\phi}} - g_{_{\theta\phi}}^2)^3} \, ,
\nonumber \\
i_6/i_1 &=& \frac{15 g_{_{\theta\phi}} \, g_{_{\phi\phi}}^2}
{(g_{_{\theta\theta}} \, g_{_{\phi\phi}} - g_{_{\theta\phi}}^2)^3} \, ,
\nonumber \\
i_7/i_1 &=& \frac{-3 g_{_{\phi\phi}}
(4 g_{_{\theta\phi}}^2 + g_{_{\theta\theta}} \, g_{_{\phi\phi}})}
{(g_{_{\theta\theta}} \, g_{_{\phi\phi}} - g_{_{\theta\phi}}^2)^3} \, ,
\nonumber \\
i_8/i_1 &=& \frac{3 g_{_{\theta\phi}}
(2 g_{_{\theta\phi}}^2 + 3 g_{_{\theta\theta}} \, g_{_{\phi\phi}})}
{(g_{_{\theta\theta}} \, g_{_{\phi\phi}} - g_{_{\theta\phi}}^2)^3} \, .
\end{eqnarray}

The ratio, which we denote by $\delta_6$,
of the integral of $\frac{1}{2} g_3^2 \exp(g_2)$
to the integral of $\exp(g_2)$ works out to be
\begin{eqnarray}
\delta_6 &=&
[2 g_{_{\theta\phi}} (3 g_{_{\theta\theta}} \, g_{_{\phi\phi}} + 2
g_{_{\theta\phi}}^2)(g_{_{\theta\theta\theta}} \, g_{_{\phi\phi\phi}} +
9 g_{_{\theta\theta\phi}} \, g_{_{\theta\phi\phi}}) \nonumber \\
&-& 3 (g_{_{\theta\theta}} \, g_{_{\phi\phi}} + 4 g_{_{\theta\phi}}^2)
(2 g_{_{\theta\theta\theta}} \, g_{_{\theta\phi\phi}} \, g_{_{\phi\phi}} +
 2 g_{_{\phi\phi\phi}} \, g_{_{\phi\theta\theta}} \, g_{_{\theta\theta}} +
 3 g_{_{\theta\theta\phi}}^2 \, g_{_{\phi\phi}} +
 3 g_{_{\phi\phi\theta}}^2 \, g_{_{\theta\theta}}) \nonumber \\
&+&30 g_{_{\theta\phi}} (g_{_{\theta\theta\theta}} \, g_{_{\theta\theta\phi}}
 \, g_{_{\phi\phi}}^2 +
g_{_{\phi\phi\phi}} \, g_{_{\phi\phi\theta}}
 \, g_{_{\theta\theta}}^2) \nonumber \\
&-&5(g_{_{\theta\theta\theta}}^2 \, g_{_{\phi\phi}}^3 +
     g_{_{\phi\phi\phi}}^2 \, g_{_{\theta\theta}}^3)] /
[24 (g_{_{\theta\theta}} \, g_{_{\phi\phi}} - g_{_{\theta\phi}}^2)^3] \, .
\label{alpha6}
\end{eqnarray}

The only remaining matter is to find the necessary values of the higher
derivatives of the function $g$ at the stationary-phase point.  The definition
of $g$, given in Eq.~(\ref{gdefinition}) can be used to find its
second derivatives, given in Eq.~(\ref{secondderivs}),
its third derivatives,
\begin{eqnarray}
\frac{\partial^3 g}{\partial \theta^3} &=&
2 (j+j_2-j_1) \csc^2 \theta \, \cot \theta
+ 2 (j_1+j_2-j) \csc^2 (\theta - \phi) \, \cot(\theta-\phi) \, , \nonumber \\
\frac{\partial^3 g}{\partial \theta^2 \, \partial \phi} &=&
-2 (j_1+j_2-j) \csc^2 (\theta - \phi) \, \cot(\theta-\phi) \, , \nonumber \\
\frac{\partial^3 g}{\partial \theta \, \partial \phi^2} &=&
2 (j_1+j_2-j) \csc^2 (\theta - \phi)\, \cot(\theta-\phi) \, , \nonumber \\
\frac{\partial^3 g}{\partial \phi^3} &=&
2 (j_1+j_2-j) \csc^2 (\theta - \phi) \, \cot(\phi-\theta)
+ 2 (j+j_1-j_2) \csc^2 \phi \, \cot \phi\, ,
\label{thirdderivs}
\end{eqnarray}
and its fourth derivatives,
\begin{eqnarray}
\frac{\partial^4 g}{\partial \theta^4} &=&
- 2 (j+j_2-j_1) \csc^2 \theta \, (3 \cot^2 \theta + 1)
\nonumber \\
&-& 2 (j_1+j_2-j) \csc^2 (\theta - \phi) [3 \cot^2(\theta - \phi) + 1],
\nonumber \\
\frac{\partial^4 g}{\partial \theta^3 \, \partial \phi} &=&
2 (j_1+j_2-j) \csc^2 (\theta - \phi) \,[3 \cot^2(\theta - \phi) + 1]
\, , \nonumber \\
\frac{\partial^4 g}{\partial \theta^2 \, \partial \phi^2} &=&
- 2 (j_1+j_2-j) \csc^2 (\theta - \phi)\,[3 \cot^2(\theta - \phi) + 1]
\, , \nonumber \\
\frac{\partial^4 g}{\partial \theta \, \partial \phi^3} &=&
2 (j_1+j_2-j) \csc^2 (\theta - \phi) \, [3 \cot^2(\theta - \phi) + 1]
\, , \nonumber \\
\frac{\partial^4 g}{\partial \phi^4} &=&
- 2 (j_1+j_2-j) \csc^2 (\theta - \phi) \, [3 \cot^2(\theta - \phi) + 1]
\nonumber \\
&-& 2 (j+j_1-j_2) \csc^2 \phi \, (3 \cot^2 \phi + 1)\, .
\label{fourthderivs}
\end{eqnarray}
Given the values of the cotangents of $\theta$ and $\phi$ at a
stationary-phase point, these derivatives can be evaluated without
having to find the angles, {\it i.e.} without having to make any
choices of branch cuts.  One way to do this is to use
Eq.~(\ref{identity19}) to evaluate $\cot(\theta - \phi)$, and the
identity $\csc^2 \theta = 1+ \cot^2 \theta$ to evaluate the squared
cosecants.

The relationship between the Clebsch-Gordan coefficient and the
integral of $\exp(g)$ is given in Eq.~(\ref{e^g-integral}).  Combining
this with our higher-order approximation for the integral results in
the following higher-order approximation for the Clebsch-Gordan
coefficient.  Each stationary-phase point contributes
\begin{equation}
(-1)^{j+m} \, (2 i)^{j+j_1+j_2} \, \pi^{-2} \,
N_{j_1 \, m_1 \, j_2 \, m_2 \, j \, m} \, 
\frac{2 \pi \, e^{g(\theta_0, \phi_0)}}{\sqrt{g_{_{\theta\theta}} \, g_{_{\phi\phi}} - g_{_{\theta\phi}}^2}} \, 
(1 + \delta_4 + \delta_6) \, .
\label{higher}
\end{equation}
As explained in the previous sections, for points in the allowed
region the sum over stationary-phase points is a sum over both of the
solutions for the cotangents of $\theta$ and $\phi$ given in
Eq.~(\ref{cotangents}), and for points in the forbidden region only
one of these solutions contributes.  Given values for $\cot\theta$ and
$\cot\phi$, Eqs.~(\ref{secondderivs}), (\ref{thirdderivs}) and
(\ref{fourthderivs}) are used to evaluate the higher derivatives of
the function $g$ at the stationary-phase point.  Then
Eqs.~(\ref{alpha4}) and (\ref{alpha6}) are used to obtain $\delta_4$
and $\delta_6$.  As shown in Eq.~(\ref{e}), the quantity
$e^{g(\theta_0, \phi_0)}$ can also be evaluated using the values of
the cotangents of $\theta$ and $\phi$.  The quantity $N_{j_1 \, m_1 \,
j_2 \, m_2 \, j \, m}$ can be approximated to sufficient accuracy
using the next correction to Stirling's approximation for the
factorials.

Finally, we present some numerical examples.  We begin with the allowed region.

For $(j_1 , m_1 , j_2 , m_2 , j , m) = (200, 100, 300, 150, 400,
250)$, the values are
\begin{eqnarray}
exact &=& 0.0703499 \, ,\nonumber \\
approx &=& 0.0703496 \, .
\end{eqnarray}

For $(j_1 , m_1 , j_2 , m_2 , j , m) = (200, 100, 300+1/2, 150+1/2,
400+1/2, 250+1/2)$, the values are
\begin{eqnarray}
exact &=& 0.0730636 \, ,\nonumber \\
approx &=& 0.0730633 \, .
\end{eqnarray}

In the forbidden region, the Clebsch-Gordan coefficients are much
smaller.  The following examples are from subregion I.

For $(j_1 , m_1 , j_2 , m_2 , j , m) = (200, 150, 300, -250, 400,
-100)$, the values are
\begin{eqnarray}
exact &=& 3.08961 \times 10^{-19} \, ,\nonumber \\
approx &=& 3.08958 \times 10^{-19} \, .
\end{eqnarray}

For $(j_1 , m_1 , j_2 , m_2 , j , m) =
(200,150,300+1/2,-250+1/2,400+1/2,-100+1/2)$, the values are
\begin{eqnarray}
exact &=& 5.32718 \times10^{-19} \, ,\nonumber \\
approx &=& 5.32712 \times10^{-19} \, .
\end{eqnarray}

Further examples of results from the higher-order approximation
are discussed in Appendix~B.

\section{conclusion}

The methods presented in this paper provide simple formulas for
calculating first-order approximations to Clebsch-Gordan coefficients
in the allowed region and in all of the forbidden subregions.
Additionally, a higher-order approximation is derived, although the
expressions are more complicated.  We do not know if the quantity
$\delta_4 + \delta_6$ in Eq.~(\ref{higher}) can be simplified when
expressed in terms of the quantum numbers (see Appendix~B for a
special case).  It appears to be complicated, as is often the case for
higher-order approximations.  The geometrical structure is not as
clear.

Our higher-order approximation provides the only known way to compute
certain digits of some Clebsch-Gordan coefficients.  By this we mean
that given any computer, we can always find quantum numbers large
enough so that the exact calculation is not feasible.  The beginning
digits may be calculated using first-order approximations; the
higher-order approximation makes it possible to compute further
digits.

The methods of this paper could also be used to derive asymptotic
expressions for the $6j$-symbols, etc.  The starting point would again
be an exact expression for the quantity of interest.  One would then
have to construct a polynomial with the property that the coefficient
of one of its terms is this exact expression.  Then an integral
expression would be obtained, and finally this integral would be
approximated using the stationary-phase method.

As mentioned in the introduction, this work could have applications in
high-angular momentum calculations and theoretical investigations
which contain sums over large numbers of Clebsch-Gordan
coefficients\cite{generatingfunctions,Labarthe}.

Our analysis in the forbidden region led us to the realization that
simple sign functions exist there that give the sign of the exact
Clebsch-Gordan coefficients.  These are summarized in Table~I.

A subject for future work is the approximation of Clebsch-Gordan
coefficients and $6j$-symbols near the boundary between the allowed
and forbidden regions.  Ponzano and Regge\cite{PR} have conjectured
and supplied numerical evidence for a typical Airy-function caustic
behavior.  Also, of course, it should be possible to extend the
present calculations to even higher orders.

\appendix

\section{A one-dimensional example}

In this appendix we consider a one-dimensional example of an integral
that gives a Fourier coefficient of a function which is an integer
power of a fixed function.  We are interested in the asymptotics
of the result for large values of the two integers involved.

We define the function $F(m,n)$ for positive integers $m$ and $n$ by
\begin{equation}
F(m,n) = \int_{-\pi/2}^{\pi/2} \cos^n x \; e^{i m x} \, dx \, .
\end{equation}
It is possible to evaluate this integral exactly in closed form:
\begin{equation}
F(m,n) = \left \{
\begin{array}{ll}
2^{-n} \pi \left ( \begin{array}{c} n \\ (n-m)/2 \end{array} \right ) &
n-m \, even \\ & \\
(-1)^{(n+1-m)/2} \, 2^{n+2} \, n! \, \frac{[(m+n+1)/2]! \, (m-n-1)!}
{(m+n+1)! \, [(m-n-1)/2]!} &
n-m \,odd, n<m \\ & \\
2^{n+2} \, n! \, \frac{[(n+1-m)/2]! \, [(n+1+m)/2]!}
{(n+1-m)! \, (n+1+m)!} &
n-m \, odd, n>m
\label{exact}
\end{array}
\right .
\end{equation}
In deriving these results, one uses the definition of the beta
function, $B(z+1, w+1) = \int_0^1 t^z (1-t)^w \, dt$, and the relation
between the beta function and the gamma function, $B(z,w) = \Gamma(z)
\Gamma(w) / \Gamma(z+w)$.  One also uses the results that for integers
$k \ge 0$,
\begin{eqnarray}
\Gamma(k + \frac{1}{2}) &=& \frac{\sqrt{\pi} \, (2k)!}{2^{2k} \, k!} \, ,
\nonumber \\
\Gamma(-k + \frac{1}{2}) &=& (-1)^k \frac{2^{2k} \, \sqrt{\pi} \, k!}{(2k)!} \, .
\end{eqnarray}

\subsection{Asymptotics of the exact expressions}

In order to compare the stationary-phase approximations derived in the
following subsection with the exact value of $F(m,n)$, we will use
Stirling's approximation for the factorials in the exact expressions
in Eq.~(\ref{exact}).  The accuracy to which we will work is that the
ratio of the exact value to the approximation should go to unity as
$n$ and $m$ go to infinity, holding the ratio of $n$ to $m$ fixed.
The difference between the logarithm of the exact expression and the
logarithm of the approximation thus goes to zero as the two integers
get large (the errors are of order $1/n$).

Stirling's approximation, through order unity (for the
logarithms), is
\begin{equation}
x! \, \approx \, \sqrt{2\pi x} \, x^x \, e^{-x}\;.
\label{Stirling}
\end{equation}
The next correction to this is a multiplicative factor of $e^{1/(12 \,
x)}$.  Thus, the ratio of $x!$ to the approximation given in
Eq.~(\ref{Stirling}) approaches unity as $x$ goes to infinity.

Our approximation of the exact expression for $F(m,n)$ works out to be
\begin{equation}
F(m,n) \approx \left \{
\begin{array}{ll}
\sqrt{\frac{2 \pi}{n}} \left( \frac{1-\frac{m}{n}}{1+\frac{m}{n}}
\right)^{m/2} \left[ 1 - \left( \frac{m}{n} \right)^2 \right]^{-(n+1)/2}
& n > m \\ & \\
0 & n-m \, even , n<m \\ & \\
(-1)^{(n+1-m)/2} \, 2 
\sqrt{\frac{2 \pi}{n}} \left( \frac{\frac{m}{n}-1}{\frac{m}{n}+1}
\right)^{m/2} \left[ \left( \frac{m}{n} \right)^2 - 1\right]^{-(n+1)/2}
& n-m \, odd, n<m
\end{array}
\right .
\label{approx exact}
\end{equation}
In deriving this result, we have used the fact that the inequality
$n>m$ implies $n - m >> 1$.  This is true because we are holding the
ratio of the two integers fixed while letting them become large.  In
other words, errors of order $1/n$ are the same order as errors of
order $1/(n-m)$.  Thus the Stirling approximation is used for
quantities such as $(n-m)!$.  Similar remarks apply to the inequality
$n<m$.

In Eq.~(\ref{approx exact}), only positive quantities are raised to
powers that could be non-integer.  Thus there are no phase
ambiguities.  If one is sloppy about phases, the last expression
appears to be the same as the first, differing only be a factor of
two.  The origin of this factor of two has a simple interpretation in
the stationary-phase approximation, described in the next subsection.

It is remarkable that the first expression (for the case $n>m$, $n-m$
even) and the third expression (for the case $n>m$, $n-m$ odd) in
Eq.~(\ref{exact}) have the same asymptotics to the order at which we
are working.  A calculation is involved in showing this.  The result
that comes from applying the Stirling approximation to the third
expression is
\begin{displaymath}
\sqrt{\frac{2 \pi}{n}} \left( \frac{1-\frac{m}{n+1}}{1+\frac{m}{n+1}}
\right)^{m/2} \left[ 1 - \left( \frac{m}{n+1} \right)^2 \right] ^{-(n+1)/2} \, .
\end{displaymath}
To the accuracy to which we are working, this turns out to be the same
as the first expression in Eq.~(\ref{approx exact}), although some
work is required to show this.

\subsection{Stationary-phase approximation}

To do a stationary-phase approximation for the function $F(m,n)$, we
write the function as
\begin{equation}
F(m,n) = \int_{-\pi/2}^{\pi/2} e^{n \, g(x)} \, dx \, ,
\label{g integral}
\end{equation}
where the function $g$ is defined by
\begin{equation}
g(z) =  \ln \cos z \, + \, i \frac{m}{n} z \, .
\end{equation}
With the usual choice of branch cut for the logarithm function, the function
$g(z)$ is analytic everywhere in the complex plane except for vertical lines
that intersect the real axis at odd multiples of $\pi$, and at the
intervals on the
real axis where $\cos z$ is nonpositive.  The identity
\begin{equation}
\cos(x+iy) = \cos x \, \cosh y - i \sin x \, \sinh y
\label{cos}
\end{equation}
is useful in showing
this.  Knowledge of the region of analyticity of $g$ allows us to deform
the contour of integration in Eq.~(\ref{g integral}) without changing the
value of the integral.  We would like to deform the contour so that the
phase of the integrand $e^{n \, g(z)}$ is constant.  To do this, we need to
know the imaginary part of $g(z)$.  With the help of Eq.~(\ref{cos})
we find that this is
\begin{equation}
\Im [ g(x+iy) ] = - \tan^{-1}( \tan x \, \tanh y) + \frac{m}{n} x \, .
\label{Im}
\end{equation}
If a contour is selected in such a way that this function is a constant,
then $g(x+iy)$ will equal an imaginary constant plus
a real-valued function along the contour.  The integrand $e^{n \, g(z)}$
will then equal a constant phase factor raised to the $n$-th power times
a fixed real-valued function raised to the $n$-th power.  This fixed
real-valued function may be approximated by a
Gaussian, and the integral may then be evaluated.
Stationary-phase points $z_s$ satisfy
the condition
\begin{equation}
g^\prime(z_s) = 0 \, .
\end{equation}
This is equivalent to the condition
\begin{equation}
\tan z_s = i \frac{m}{n} \, .
\end{equation}
We note that the value of $g^{\prime\prime}$ at a stationary-phase
point is
\begin{equation}
g^{\prime\prime}(z_s) = -\sec^2 \, z_s = \left(\frac{m}{n}\right)^2 - 1 \, .
\end{equation}

It is necessary to distinguish two cases, the case $m<n$ and the
case $m>n$ (recall that $m$ and $n$ are both positive by assumption).
We first consider the case $m<n$.  In this case, it follows from
a study of Eq.~(\ref{Im}) that a constant phase contour exists
that connects the endpoints of the integral and passes through
the stationary-phase point
\begin{equation}
z_s = i \, \tanh^{-1} \frac{m}{n}
\end{equation}
in a direction that is parallel to the real axis.  The integrand is
approximated by
\begin{displaymath}
e^{n \, g(z_s) + n \, g^{\prime\prime}(z_s)(z-z_s)^2/2} \, ,
\end{displaymath}
and the result for the integral is
\begin{equation}
\sqrt{\frac{2 \pi}{-n \, g^{\prime\prime}(z_s)}} \, e^{n \, g(z_s)}
= \sqrt{\frac{2 \pi}{n}} \left( \frac{1-\frac{m}{n}}{1+\frac{m}{n}}
\right)^{m/2} \left[ 1 - \left( \frac{m}{n} \right)^2 \right]^{-(n+1)/2} \, ,
\end{equation}
which agrees with the result in Eq.~(\ref{approx exact}).

We now move on to the case $m>n$.  In this case, no single contour
exists with the properties that it connect the endpoints of the integral
and that the quantity in Eq.~(\ref{Im}) be constant.  Instead, we choose
a contour consisting of three straight-line pieces.   The first part, $C_1$,
is defined to start at $-\pi/2$ and go vertically upwards to $-\pi/2 + i Y$,
$Y$ being a large positive
real number.  The second part, $C_2$, is defined to go
from $-\pi/2 + i Y$ to $\pi/2 + i Y$, and the third part, $C_3$, goes straight
down to the $\pi/2$ endpoint of the integral.  The parts $C_1$ and $C_3$
contain stationary-phase points, which we call $z_{s-}$ and $z_{s+}$, and
which are given by
\begin{equation}
z_{s\pm} = \pm \frac{\pi}{2} + i \, \tanh^{-1} \frac{n}{m} \, .
\end{equation}
For the integral along $C_1$ the integrand is approximated by
\begin{displaymath}
e^{n \, g(z_{s-}) + n \, g^{\prime\prime}(z_{s-})(z-z_{s-})^2/2} \, .
\end{displaymath}
We parametrize the curve $C_1$ by $z = z_{s-} + i \, t$, where $t$ is
a real parameter.  Then $dz$ is $i \, dt$ and the resulting
approximation for the integral along $C_1$ is
\begin{equation}
i \, \sqrt{\frac{2 \pi}{+n \, g^{\prime\prime}(z_{s-})}} \, e^{n \, g(z_{s-})} \, .
\label{C1}
\end{equation}
Similarly, the approximation for the integral along $C_3$ is
\begin{equation}
-i\,\sqrt{\frac{2 \pi}{+n \, g^{\prime\prime}(z_{s+})}} \, e^{n \, g(z_{s+})} \, .
\label{C3}
\end{equation}
Because of the condition $m>n$ the integral along the curve $C_2$ goes
to zero as $Y$ goes to infinity.  The approximation for $F(m,n)$ is
thus the sum of the expressions given in Eqs.~(\ref{C1}) and
(\ref{C3}).  If $m-n$ is odd the sum is zero, in agreement with the
exact value.  If $m-n$ is even, the sum agrees with the approximation
of the exact result, given in Eq.~(\ref{approx exact}).

Thus we see that depending on the ratio $m/n$, different numbers of
stationary-phase points must be considered due to fundamental changes
in the form of the stationary-phase contours as $m/n$ goes from one
side of the critical value of 1 to the other side.  On either side of
the critical value of 1, a stationary-phase approximation is possible.
The behavior near the critical value is discussed briefly in the main
part of this paper.

\section{The case of vanishing magnetic quantum numbers}

Clebsch-Gordan coefficients for the case of vanishing magnetic quantum
numbers ($m_1 = m_2 = m = 0$) are of interest in atomic and nuclear
physics.  Many of the expressions derived in this paper simplify in
this case.  Also, comparisons with the asymptotics of the exact
closed-form expression, given in Eq.~(\ref{exactm0}), are possible.

The vanishing of the magnetic quantum numbers implies that $\beta$,
defined in Eq.~(\ref{beta}), is real.  The case $\beta = 0$ is simple
because one of the $j$ quantum numbers is then equal to the sum of the
other two, and the integral in Eq.~(\ref{int}) may be evaluated exactly
with a small amount of effort.  Thus, we will consider the case $\beta
> 0$.  Two other facts that will be used throughout this appendix are
that the set of quantum numbers is in the allowed region (since
$\beta$ is real) and that the $j$ quantum numbers are integers (since
$j_i - m_i$ is always an integer).

First, we work out the simplifications that occur in the all-real
expression in Eq.~(\ref{ourexprsimp}).  Equation~(\ref{chi}) becomes
\begin{equation}
\chi = \frac{\pi}{2}\left(j + j_1 + j_2 + \frac{3}{2} \right) 
\label{chi0}
\end{equation}
and Eq.~(\ref{ourexprsimp}) becomes
\begin{equation}
\langle j_1\,0\,j_2\,0 \,|\,j\,0 \rangle \approx
2  \, \sqrt{\frac{j}{\pi \beta}} \, \cos
\left[\frac{\pi}{2} (j_1 + j_2 - j) \right] \, .
\end{equation}
This agrees with the first-order approximation of the exact
expression, which can be obtained from the higher-order approximation
[Eq.~(\ref{exactm0approx})] of the exact result, given in
Eq.~(\ref{exactm0}).  If $j_1 + j_2 - j$ is odd, then both of the
expressions are zero.  (In this case $j_1 + j_2 + j$ is also odd since
$2j$ is even.)  If $j_1 + j_2 - j$ is even, then both have a sign of
$(-1)^{(j_1+j_2-j)/2}$.

Next, we move on to the higher-order approximation.  From
Eq.~(\ref{cotangents}) it is apparent that the two solutions for the
cotangents of $\theta$ and $\phi$ are related by simply reversing the
signs.  It follows from Eqs.~(\ref{secondderivs}), (\ref{thirdderivs})
and (\ref{fourthderivs}) that the values of the second and
fourth-order derivatives of $g(\theta, \phi)$ are unchanged, while the
third-order derivatives have their signs flipped.
Equations~(\ref{alpha4}) and (\ref{alpha6}) imply that the quantities
$\delta_4$ and $\delta_6$ are unchanged.  Finally, Eq.~(\ref{e})
implies that the quantity $e^{g(\theta_0, \phi_0)}$ gets multiplied by
$(-1)^{j_1 + j_2 - j}$ for the second root.  Because $j$ is an
integer, this phase factor is the same as $(-1)^{j_1 + j_2 + j}$.
Since we are in the allowed region, we must sum over both
stationary-phase points.  We see that for odd values of ${j_1 + j_2 +
j}$ the result is zero, while for even values of ${j_1 + j_2 + j}$ the
result is twice the contribution obtained from one of the
stationary-phase points.

The values of the second derivatives of $g(\theta, \phi)$ at the
stationary-phase points are obtained from Eqs.~(\ref{cotangents}) and
(\ref{secondderivs}), and the results simplify quite a bit.
\begin{eqnarray}
g_{_{\theta\theta}} &=&
  - {{4\,{j_2}\,\left( {j_1} + j \right) }\over 
    {{j_1} + {j_2} + j}} \, , \\
g_{_{\theta\phi}} &=&
  {{4\,{j_1}\,{j_2}}\over {{j_1} + {j_2} + j}} \, , \\
g_{_{\phi\phi}} &=&
  - {{4\,{j_1}\,\left( {j_2} + j \right) }\over 
    {{j_1} + {j_2} + j}}\, .
\end{eqnarray}
From these equations results the following expression for the
determinant of the $2\times2$ Hessian matrix of second derivatives of
$g(\theta, \phi)$.
\begin{equation}
\det \frac{\partial^2 g}{\partial (\theta, \phi)^2} =
  {{16\,{j_1}\,{j_2}\,j}\over {{j_1} + {j_2} + j}}
\label{number6}
\end{equation}
We note that this result is nonzero, and it does not vanish in any
special cases that have nonzero $j$ quantum numbers, which seems to
contradict the statement made at the end of Sec.~\ref{secIII} about the
vanishing of the determinant when $\beta$ vanishes.  The resolution of
this apparent contradiction has to do with the fact that in the
stationary-phase analysis of this paper we do not simultaneously
consider the cases $m_1 = m_2 = m = 0$ and $\beta = 0$.  These two
conditions together would imply $\alpha$ is zero.  The quantity
$\alpha$ is defined in Eq.~(\ref{alpha-lambda}).  It vanishes when
$j=j_1+j_2$, $j_1=j_2+j$ or $j_2=j+j_1$.  In general, as long as
$\alpha$ is nonzero, the expression for the determinant can be put
(after some work) in the form
\begin{equation}
\det \frac{\partial^2 g}{\partial (\theta, \phi)^2} =
\frac{\beta \, P_1 + \beta^2 \, P_2}{\alpha^2 (j+j_1+j_2)^2} \, ,
\label{detbeta}
\end{equation}
where $P_1$ and $P_2$ are (large) polynomials in the quantum numbers.
This equation justifies the statement that the determinant is zero in
cases where $\beta$ is zero.  On the other hand, in cases where $m_1 =
m_2 = m = 0$, $P_1$ vanishes and $\alpha$ and $\beta$ are equal, and
the result simplifies to that shown in Eq.~(\ref{number6}).  The case
of $m_1 = m_2 = m = 0$ and $\beta = 0$ requires a separate treatment.
It is necessary to go back to the original integral representation for
the Clebsch-Gordan coefficient.  The integral may be approximated by
the methods of stationary phase, but it is simpler just to evaluate it
or Eq.~(\ref{label215}) exactly, which is possible at that point.

Higher-order derivatives of $g(\theta, \phi)$ at the stationary-phase
points simplify as well.  Equations~(\ref{alpha4}) and (\ref{alpha6})
for $\delta_4$ and $\delta_6$ yield results that are much simpler than
for the general case of nonzero magnetic quantum numbers.
\begin{eqnarray}
\delta_4 + \delta_6 &=& ( {{{j_1}}^5}\,{j_2}  -
      2\,{{{j_1}}^3}\,{{{j_2}}^3} + {j_1}\,{{{j_2}}^5} +
      {{{j_1}}^5}\,j - {{{j_1}}^3}\,{{{j_2}}^2}\,j -
      {{{j_1}}^2}\,{{{j_2}}^3}\,j + {{{j_2}}^5}\,j -
      {{{j_1}}^3}\,{j_2}\,{{j}^2} -
     10\,{{{j_1}}^2}\,{{{j_2}}^2}\,{{j}^2} \nonumber \\
&-&   {j_1}\,{{{j_2}}^3}\,{{j}^2} -
      2\,{{{j_1}}^3}\,{{j}^3} -
      {{{j_1}}^2}\,{j_2}\,{{j}^3} -
      {j_1}\,{{{j_2}}^2}\,{{j}^3} - 
      2\,{{{j_2}}^3}\,{{j}^3} + {j_1}\,{{j}^5} +
      {j_2}\,{{j}^5}) / ( {12\,j\,{j_1}\,{j_2}\, \beta^2 } )
\label{delta4delta6}
\end{eqnarray}
This expression may be rewritten in a more compact form, as explained
after Eq.~(\ref{exactm0approx}).  As discussed above, for odd values
of $j + j_1 + j_2$ the higher-order approximation of the
Clebsch-Gordan coefficient vanishes identically.  For even values of
$j + j_1 + j_2$, the result reduces to
\begin{eqnarray}
\langle j_1\,0\,j_2\,0 \,|\,j\,0 \rangle &\approx&
2 \, (-1)^\frac{j_1+j_2-j}{2} \sqrt{\frac{2j+1}{2\pi \beta}}
\sqrt\frac{j + j_1 + j_2}{j+j_1+j_2+1} \, (1+ \delta_4 + \delta_6 )
\left[1 + \frac{1}{24} \left(
\frac{2}{j} \right .  \right . \nonumber \\
&+& \left . \left .
\frac{2}{j_1} +\frac{2}{j_2}
-\frac{1}{j+j_1+j_2}-\frac{1}{-j+j_1+j_2}-\frac{1}{j-j_1+j_2}
-\frac{1}{j+j_1-j_2} \right) \right] \, ,
\label{higher2}
\end{eqnarray}
where we have approximated the factorials in
$N_{j_1 \, m_1 \, j_2 \, m_2 \, j \, m}$ using
the form of Stirling's approximation that
is appropriate for this order,
\begin{equation}
x! \, \approx \, \sqrt{2\pi x} \, x^x \, e^{-x} \, \left(1+\frac{1}{12\,x}
\right) \, .
\label{Stirling2}
\end{equation}

The exact value of the Clebsch-Gordan coefficient is (Ref.~\onlinecite{BL},
p. 87; Ref.~\onlinecite{Racah})
\begin{equation}
\langle j_1\,0\,j_2\,0 \,|\,j\,0 \rangle =
\left\{ \begin{array}{ll} 0 & j+j_1+j_2 \,\, {\rm odd} \\
(-1)^\frac{j_1+j_2-j}{2} \sqrt{2j+1}
\frac{\sqrt{
\frac{(-j_1+j_2+j)! (j_1-j_2+j)! (j_1+j_2-j)!}{ (j_1+j_2+j+1)!}} }
{{
\left(\frac{-j_1+j_2+j}{2}\right) ! \,
\left(\frac{j_1-j_2+j}{2}\right) ! \,
\left(\frac{j_1+j_2-j}{2}\right) ! 
 }/{
\left(\frac{j_1+j_2+j}{2}\right) !
}} &
j+j_1+j_2 \,\, {\rm even} \end{array} \right .
\label{exactm0}
\end{equation}
and this may be approximated using Eq.~(\ref{Stirling2}).  The result is
\begin{equation}
\langle j_1\,0\,j_2\,0 \,|\,j\,0 \rangle \approx
\left\{ \begin{array}{ll} 0 & j+j_1+j_2 \,\, {\rm odd} \\
2\, (-1)^\frac{j_1+j_2-j}{2} \sqrt{\frac{2j+1}{2\pi \beta}}
\sqrt\frac{j + j_1 + j_2}{j+j_1+j_2+1}
\left(1 - \frac{j j_1 j_2}{\beta^2} \right)   &
 j+j_1+j_2 \,\, {\rm even} \end{array} \right .
\label{exactm0approx}
\end{equation}

To the order that we are working, the higher-order stationary-phase
result, given in Eq.~(\ref{higher2}), and the corresponding
approximation of the exact result, given in Eq.~(\ref{exactm0approx}),
agree.  Equation~(\ref{higher2}) contains two factors that have the
form of unity plus a small correction.  If these are multiplied out
and only the first-order terms are kept, the result is the factor of
$\left(1 - \frac{j j_1 j_2}{\beta^2} \right)$ in
Eq.~(\ref{exactm0approx}).  This shows that Eqs.~(\ref{higher2}) and
(\ref{exactm0approx}) are equivalent, and it also provides an
alternative way of writing the expression in Eq.~(\ref{delta4delta6}).

\newpage

\begin{tabbing}
\indent \= forbidden subregion
\qquad \= choice of root
\qquad \= sign function
\qquad \=\kill
\>  \> $\; \;  {\rm Table \;I}$\>  \>  \\ [6pt]
\> forbidden subregion \> choice of root \> sign function \> largest $\lambda$ \\ [6pt]
\> {\rm I} \> {\sl lower} \> $1$ \> $\lambda_3$ \\
\> {\rm II} \> {\sl upper} \> $(-1)^{j_1-m_1}$ \> $\lambda_2$ \\
\> {\rm III} \> {\sl lower} \> $(-1)^{j_1-j+m_2}$ \> $\lambda_1$ \\
\> {\rm IV} \> {\sl upper} \> $(-1)^{j_1+j_2-j}$ \> $\lambda_3$ \\
\> {\rm V} \> {\sl lower} \> $(-1)^{j_2-j-m_1}$ \> $\lambda_2$ \\
\> {\rm VI} \> {\sl upper} \> $(-1)^{j_2+m_2}$ \> $\lambda_1$
\end{tabbing}

\vspace{1in}
Table I:  For each forbidden subregion, the choice of root in Eq.~(\ref{cotangents}),
the sign function as in Eq.~(3.30) and the largest $\lambda$ [which
determines the form of $\chi$, as in Eq.~(3.31)] are given.

\newpage

Figure Captions.

\vspace{.3in}

Fig.~1:  An example of a choice of three ${\bf j}$-vectors, demonstrating
that a set of quantum numbers is classically allowed.

\vspace{.3in}

Fig.~2:  The triangle-allowed region and the classically allowed region,
shown in the $m_1$-$m_2$ plane, for the case of $j$-values in the ratio
$j:j_1:j_2 = 4:2:3$.  The six forbidden subregions are labeled with
Roman numerals.


\begin{thebibliography}{99}

\bibitem{Wigner}Eugene P. Wigner, {\it Group Theory and its
Application to the Quantum Mechanics of Atomic Spectra} (Academic
Press, New York, 1959).

\bibitem{edmonds}A.R. Edmonds, {\it Angular Momentum in Quantum
Mechanics} (Princeton University Press, Princeton, NJ, 1968).

\bibitem{bandt}P.J. Brussaard and H.A. Tolhoek, Physica {\bf 23},
955(1957).

\bibitem{PR}G. Ponzano and T. Regge in {\it Spectroscopic and group
theoretical methods in physics} (North-Holland Publ. Co., Amsterdam,
1968).

\bibitem{BM}William H. Miller in {\it Advances in Chemical Physics,
Vol. 25} (Wiley, New York, 1974) edited by I. Prigonine and
S. A. Rice.

\bibitem{RR}K. Srinivasa Rao and V. Rajeswari, {\it Quantum Theory of
Angular Momentum, Selected Topics} (Springer/Narosa, Berlin/New Delhi,
1993).

\bibitem{BL}Biedenharn, L. C. and J. D. Louck, {\it Angular momentum
in quantum physics: Theory and application} (Addison-Wesley Pub. Co.,
Reading, Mass., 1981).

\bibitem{generatingfunctions}Schnetz, Oliver, ``Generating Functions for
Multi-$j$-Symbols'', LANL preprint math-ph/9805027(1998).

\bibitem{Labarthe}J.-J. Labarthe, J. Phys.\ A, {\bf 8}, 1543(1975).

\bibitem{Regge}T. Regge, Nuovo Cimento {\bf 10}, 544(1958).

\bibitem{Racah}G. Racah, Phys.\ Rev.\ {\bf 62}, 438(1942); Phys.\
Rev.\ {\bf 63}, 367(1943).

\end{thebibliography}
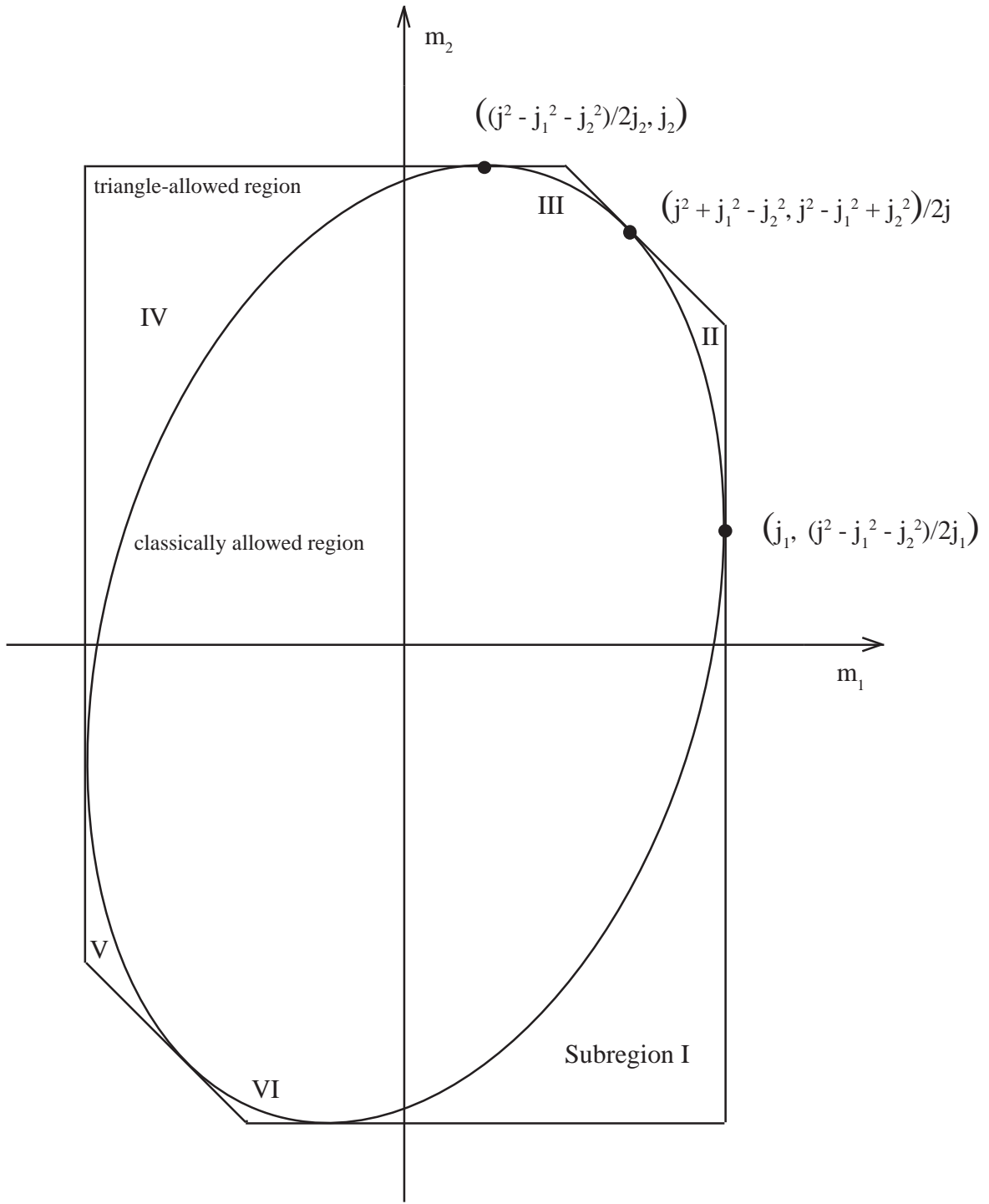
\end{document}